# Biogeochemistry-Informed Neural Network (BINN v1.0) for Improving Accuracy of Model Prediction and Scientific Understanding of Soil Organic Carbon Storage

Haodi Xu[1, †], Joshua Fan[2, †], Feng Tao[3,4 †], Lifen Jiang[1], Fengqi You[5], Benjamin Houlton[3], Ying Sun[1], Carla Gomes[2], and Yiqi Luo[1*]

[1]Soil and Crop Sciences Section, School of Integrative Plant Science, Cornell University, Ithaca, New York, USA

[2]Department of Computer Science, Cornell University, Ithaca, New York, USA

[3]Department of Ecology and Evolutionary Biology, Cornell University, Ithaca, New York, USA

[4]Currently at: Department of Informatics and Intelligent Systems, Institute of Energy and the Environment, The Pennsylvania State University

[5]Department of Systems Engineering, Cornell University, Ithaca, New York, USA

[†]These authors contributed equally to this work.

[*]Correspondence to: Yiqi Luo (yiqi.luo@cornell.edu), Feng Tao (feng.tao@psu.edu), Carla Gomes (gomes@cs.cornell.edu)

**Abstract**

The increasing availability of large-scale observational data and the rapid development of artificial intelligence (AI) provide unprecedented opportunities to enhance our understanding of the global carbon cycle and other biogeochemical processes. However, retrieving mechanistic knowledge from these large-scale data remains a challenge. Here, we develop a Biogeochemistry-Informed Neural Network (BINN) that seamlessly integrates a vectorized process-based soil carbon cycle model (i.e., Community Land Model version 5, CLM5) into a neural network (NN) structure to examine mechanisms governing soil organic carbon (SOC) storage from big data. BINN demonstrates high accuracy in retrieving biogeochemical parameter values from synthetic data in a parameter recovery experiment. Furthermore, by incorporating Monte Carlo (MC) dropout to generate posterior distributions, we demonstrate that BINN can effectively quantify uncertainty in estimated parameters. We use BINN to predict six major processes (or components in process-based models) regulating the soil carbon cycle from 25,925 observed SOC profiles across the contiguous US and compare them with the same processes previously retrieved by a Bayesian inference-based PROcess-guided deep learning and DAta-driven modeling (PRODA) approach. The good agreement between the spatial patterns retrieved by BINN and PRODA (average correlation coefficient = 0.86) suggests that BINN's ability of capturing mechanistic knowledge is consistent with the established Bayesian-based methods. Additionally, the integration of neural networks and process-based models in BINN improves computational efficiency by more than 50 times over PRODA. We conclude that BINN is an efficient framework that harnesses the power of both AI, large-scale data, and process-based modeling to understand large scale soil carbon cycle. Applying BINN to other domains of Earth system science will facilitate new scientific discoveries while improving the predictive performance of Earth system models.

**Plain Language Summary**

Conventional artificial intelligence (AI) is great at predicting biogeochemical patterns (such as the amount of organic carbon in soils) across space and time, when plenty of data are available. However, it is difficult to use these AI tools to understand the underlying mechanisms that drive these patterns, because many of these processes cannot be directly measured. In this study, we introduce a new framework called Biogeochemistry-Informed Neural Network (BINN) that integrates a process-based model into a multilayer neural network. By integrating large-scale observational datasets, BINN optimizes process-based model predictions while retrieving critical processes governing SOC storage within the framework of the chosen process-based model. Critically, BINN is more than 50 times faster than existing approaches that use Bayesian inference to align scientific models with observational data. This advance in understanding soil carbon cycling and computational speed could help scientists better predict the soil carbon cycle under different environments. BINN can be applied to retrieve mechanistic understanding of other issues of Earth's biogeochemical cycles from large-scale data with process-based models.

## 1. Introduction

Artificial intelligence (AI) has revolutionized our ability to leverage big data to uncover relationships in complex systems such as the Earth system[1]. Machine learning and deep learning, for example, have shown power in discovering key patterns from data in biogeochemistry, such as representing soil organic carbon (SOC) concentrations[2,3], predicting aboveground carbon accumulation rates in naturally regenerating forests[4], and estimating soil respiration[5]. However, most AI-based approaches primarily learn black-box statistical correlations from the data rather than causality, making it challenging to translate the learned relationships and patterns into mechanisms and controls on processes. This lack of mechanistic insight is an inherent weakness of AI-based models[6].

To address this challenge, various hybrid approaches have emerged, aiming to integrate scientific knowledge and reasoning with standard machine learning methods. This integration leverages the strengths of data-driven techniques alongside scientific theory[7,8]. By introducing physical or knowledge-based constraints and reasoning, AI-based predictions can be further constrained by scientific knowledge in addition to being driven by observational data. More importantly, unlike the uninterpretable weights and biases in a conventional neural network (NN), the latent physical parameters embedded in the knowledge-based constraints explicitly represent physical and biological processes, providing mechanistic interpretability to the neural network's predictions. For example, we can use physical knowledge to regularize the model towards physically-consistent predictions[9], such as by adding loss terms that penalize violations of a known governing equation in physics-informed neural networks (PINN)[10], or discrepancies between neural network and process-based model outputs[11]. These methods do not guarantee that NN simulations will strictly obey physical principles; they only softly discourage violations.

While recent 'Hard-PINNs' developments have sought to enforce these constraints more strictly to minimize the difference between NN predictions and numerical model simulations[12], this requires a complex multi-level optimization procedure with many hyperparameters, and still does not satisfy the constraint exactly. Another body of work thus tries to embed physical knowledge inside the NN model itself, such as modifying an Long Short-Term Memory (LSTM) network to guarantee mass conservation at each step[13], or embedding hydrological models[14,15], terrestrial models[16], thermodynamic rules[17], domain-specific objectives[18], or Bragg's law for x-ray diffraction[19] into the NN model architecture. Recent studies have further advanced these end-to-end setup to infer stomatal resistance and aerodynamic resistance in evapotranspiration modeling[20] and to estimate global physical parameters for water cycle models[21]. Despite these advances, previous efforts mostly used limited constraints in a system (e.g., a few empirical relationships in photosynthesis[22]) or simplified models that comprise latent variables and conservation principles[14]. It remains challenging to integrate numerous differential equations into a NN to study the dynamics of a complex system, such as SOC dynamics. This difficulty stems from the tremendous computational cost required to execute complex models, and the associated high-dimensional parameter spaces that need to be optimized for predictive model simulations. In particular, the capacity of NN to efficiently optimize process-related parameter values with faithful identifiability remains an under-explored area of research.

Meanwhile, a traditional method that uses scientific knowledge to guide inference of biogeochemical processes from observational data is data assimilation. This method has played a central role in advancing our understanding of many processes, such as terrestrial carbon flux partitioning[9], nutrient cycling[23], ecosystem respiration[1], plant phenology[24], and hydrology–biogeochemistry couplings[25]. However, the data assimilation method usually estimates a set of fixed biogeochemical parameters without spatial or temporal variations, which has been argued not represent variations in ecosystem properties over space and time[26]. While statistical methods such as Geographically-Weighted Regression allow spatially-varying regression coefficients, they can only model linear relationships between inputs and outputs at each location [27,28]. A recently-developed approach, PROcess-guided deep learning and DAta driven modeling (PRODA), first uses Bayesian inference-based data assimilation methods to find the optimal biogeochemical parameters for each site, and then trains a deep learning model to infer biogeochemical parameters at new sites, improving our understanding of the global soil carbon cycle[29–32]. This approach harnesses the strengths of both process-based modeling and deep learning methods to improve SOC simulations by allowing biogeochemical parameters to vary across space. However, the Bayesian inference-based, site-level data assimilation used in the PRODA approach requires vast computational resources, making the method inefficient in both time and energy and thus difficult to apply broadly.

In this study, we integrate a matrix form of a process-based model (i.e., Community Land Model version 5, CLM5) that describes SOC dynamics into a neural network, thus developing a Biogeochemistry-Informed Neural Network (BINN). BINN is an end-to-end framework that combines data-driven machine learning with process-based modeling to facilitate the interpretability of model-represented biogeochemical dynamics, such as SOC dynamics in this study. Herein, we first introduce the structure of BINN, followed by a demonstration of BINN's ability to recover biogeochemical parameters with high accuracy through a parameter recovery experiment. We further implemented Monte Carlo (MC) dropout to generate posterior parameter distributions to quantify uncertainties and identify equifinality issues. Lastly, we evaluate model components from real-world SOC observations. The BINN-predicted biogeochemical parameters accurately simulate SOC observations and are similar to those produced by PRODA, while being much faster to compute. By combining data-driven learning with the mechanistic representations of process-based models, BINN can accurately and quickly infer underlying physical processes from SOC observations.

## 2. Methods

### 2.1 Biogeochemistry-Informed Neural Network (BINN)

BINN integrates a process-based model (CLM5, see Section 2.1.2 below) into a differentiable, end-to-end framework (Fig. 1a) that maps environmental covariates to site-level biogeochemical parameters, feeds those parameters into CLM5 for the forward simulation, and back-propagates the loss through the whole framework during training (Fig. 1b).

We implemented BINN in Python using PyTorch and executed the experiments on the supercomputer, Derecho, maintained by National Center for Atmospheric Research (NCAR). The experiments were conducted using one compute node with 128 CPU cores, leveraging PyTorch Distributed Data Parallel (DDP) for multi-CPU training. Meanwhile, BINN was also tested with a multi-GPU compute node on a cluster in Cornell University's Center for Advanced Computing, confirming its capability for training on GPU clusters when needed.

#### 2.1.1 Neural Network

A fully-connected neural network learns the relationship between environmental covariates (Table S1) and biogeochemical parameters (Table S2) over space. At each site, we have 60 environmental covariates (input features). While the numeric features can be directly passed into the neural network, we use embedding layers to encode categorical (discrete) covariates and spatial coordinates. For each categorical covariate, the embedding layer learns a vector embedding for each category (e.g. climate type). For spatial coordinates, we pass the latitude and longitude through a spatial positional encoder to obtain a location embedding vector, characterizing unobserved aspects of each location that are not captured in our covariates[33].

We then concatenate the categorical covariate embeddings, location embedding, and the remaining covariates into a vector $\boldsymbol{e}$, and pass this through a 4-layer neural network $f_{NN}$ (Equation 1) with learnable weights/biases $\boldsymbol{w}$, which outputs a vector $\widetilde{\boldsymbol{p}}$ with 21 values (representing *unconstrained* predictions for each parameter in the process-based model):

$$\widetilde{\boldsymbol{p}} = f_{NN}(\boldsymbol{e}; \boldsymbol{w}) \tag{1}$$

Each layer of the neural network comprises prescribed numbers of neurons that receive information either from the environmental covariates (for the first layer) or outputs of the previous layer. It then computes linear transformations of the input:

$$y = \sum w_i x_i + b \tag{2}$$

where $y$ is an output from a neuron after the linear combination of its input, $w_i$ is a learnable weight for $x_i$, $x_i$ is the input from either the environmental covariates or the previous layer's outputs, and $b$ is a learnable neuron-specific bias. After the linear transformation, a nonlinear activation function is applied to generate the eventual results at each neuron, so that the neural network can generate complex nonlinear relationships between the inputs (i.e., environmental covariates) and the outputs (i.e., the biogeochemical parameters in the process-based model). Note that the linear transformations and activation functions are both differentiable, so we can backpropagate gradients through the network. This allows us to calculate the gradient of the loss function with respect to all learnable weights $w_i$ and biases $b$. To prevent overfitting and enable uncertainty quantification through MC dropout, we incorporated a dropout layer after each hidden layer (Equation 2). During BINN training process, neurons are randomly deactivated with a prescribed dropout rate (see below for hyperparameter selections), allowing the network to explore a range of model configurations.

In our study, for the first three layers of the neural network, we assign each of them to have 128 neurons to process information from the previous layer and use *LeakyReLU* as the activation function:

$$LeakyReLU(y) = max(0, y) + negative_slope * min(0, y) \tag{3}$$

where *max(0, y)* is a function that returns the larger value of *0* or *y*, while *min(0, y)* returns the smaller value of *0* or *y*. The *negative_slope* is a hyperparameter that determines how "leaky" the function is for negative *y*. *LeakyReLU* is chosen over traditional *ReLU* because it allows gradients to flow through the network even when the inputs to the activation are negative.

The final layer has 21 output neurons, with one corresponding to each biogeochemical parameter in the process-based model. We do not use *LeakyReLU* after the final linear transformation because it would bias the distribution of the predicted parameters. However, as we have prior knowledge about the plausible range of values for each biogeochemical parameter, the initial parameter predictions $\tilde{\boldsymbol{p}}$ are first normalized to be in the range [0, 1] by element-wise *parameterized sigmoid* functions σ (Equation 4) and then further linearly scaled into the predicted parameters $p$ (Equation 5). This ensures that each parameter $p_i$ stays within their prior ranges:

$$p_{norm,i} = \sigma(\tilde{p}_i, \gamma\ ) = \frac{1}{1 + exp\left(-\frac{\tilde{p}_i}{\gamma}\right)} \tag{4}$$

$$p_i = f_{scale}\left(p_{norm,i}, \theta_{i,max}, \theta_{i,min}\right) = p_{norm,i} * (\theta_{i,max} - \theta_{i,min}) + \theta_{i,min} \qquad (5)$$

where $\tilde{p}_i$ is the i-th output of the neural network, $\gamma$ is a learnable parameter, and $\theta_{i,max}$ and $\theta_{i,min}$ are the plausible upper and lower limits for each biogeochemical parameter $p_i$. Thus, the final output of the neural network is $p$, a vector of 21 biogeochemical parameters for each location, with each parameter $p_i$ being constrained to be in its prior range $[\theta_{i,min}, \theta_{i,max}]$. $\theta_{i,min}$ and $\theta_{i,max}$ are values taken from previous literature to indicate plausible limits for processes quantified by each biogeochemical parameter $p_i$[31]. Note that we introduce a $\gamma$ value in Equation 4 to control how fast the results after activation function can converge to 0 or 1 and to $\theta_{i,max}$ or $\theta_{i,min}$ after scaling. When the $\gamma$ value is small, the predicted parameters may quickly get stuck at $\theta_{i,min}$ and $\theta_{i,max}$. If this happens, the derivative of the activation approaches zero and it will be difficult to further optimize the predictions via gradient-based optimization (see Section 2.1.4)[34]. Thus, we tuned $\gamma$ to be a relatively large value to facilitate neural network optimization.

We conducted a hyperparameter search to determine suitable settings for BINN (i.e., epochs of training, batch size, CPU number, optimizer, learning rate, dropout rate, embedding size of the embedding layer, whether to use batch normalization, the initialization of $\gamma$, *negative_slope* in the *LeakyReLU*, and the loss function hyperparameters). Based on a grid search for these hyperparameters, we chose to train BINN for 300 epochs with a batch size of 32, using PyTorch DDP to distribute training across 128 CPUs, as this performed best on our validation dataset. The optimized BINN model was recorded each time when the validation loss improved over the previous best model. We used the *AdamW* optimizer with a learning rate of 0.01. The embedding size was set as 64, and the dropout rate was set as 0.3. The network also used batch normalization after each *LeakyReLU* activation function to normalize the layers' outputs by re-centering and re-scaling, making training faster and more stable[35]. We initialized $\gamma$ to 64.5 but allowed it to be further optimized throughout the training processes within the range from 10 to 109. Specifically, we used a sigmoid to constrain the range:

$$\gamma = 10 + 109 \cdot \frac{1}{1 + exp(-\gamma')} \tag{6}$$

where $\gamma'$ is a learnable parameter that is initialized to 0. We set *negative_slope* to -0.3 in the *LeakyReLU* by default.

We also performed a quantitative sensitivity analysis to assess how BINN performance responds to variations in the loss and activation functions (Table S3). Specifically, we tested the sensitivity of replacing the smooth L1 loss with L1 loss and L2 loss and substituting the hyperbolic cosine parameter loss with a squared loss. Furthermore, we evaluated the sensitivity to changes in activation functions by modifying the first three layers from leaky ReLU to ReLU and the final layer from Sigmoid to hard Sigmoid.

#### 2.1.2 Process-based Model

In this study, we use a matrix form of the soil carbon module of CLM5 to represent our knowledge of SOC dynamics[36,37] (Fig. S1). CLM5 has been continuously developed and refined over the past decade for simulating biogeochemical cycles, including SOC dynamics[38]; it mathematically represents our knowledge of SOC dynamics with 140 partial differential equations. We chose CLM5 to enable direct comparison with PRODA, which uses the same model.

CLM5 simulates SOC dynamics across 20 soil layers to 8 m. Each layer contains 7 carbon pools, including one coarse woody debris pool, three litter pools corresponding to metabolic, cellulose, and lignin materials, and three SOC pools classified into fast, slow, and passive pools according to their turnover times. This structure results in a total of 140 carbon pools (7 pools × 20 layers).

A key innovation of our approach is that we incorporate into our neural network framework a differentiable CLM5 model, whose structure can be represented in a matrix form[37,39] as:

$$\frac{dX(t)}{dt} = B(t)I(t) - A\xi(t)KX(t) - V(t)X(t) \quad (7)$$

where $dX(t)/dt$ is carbon pool change at time *t*; $I(t)$ is the total carbon input from vegetation at time *t*, $B(t)$ (140×1) is the allocation of carbon input to different pools; $A$ (140×140) is the carbon transfer matrix, quantifying horizontal carbon movement between pools in the same layer; $K$ (140×140) is the intrinsic decomposition rate of each carbon pool, which is the same for each pool across 20 layers; $\xi(t)$ (140×140) captures how the environment modifies the intrinsic decomposition rate in the $K$ matrix by temperature ($\xi_T$), water ($\xi_W$), oxygen ($\xi_O$), and depth ($\xi_D$) scalars; $V(t)$ (140×140) defines how SOC enters and leaves each layer; and $X(t)$ is carbon pool size. The term $B(t)I(t)$ represents the vegetation carbon input, $A\xi(t)KX(t)$ describes the SOC movements among the 7 pools within each layer, and $V(t)X(t)$ indicates vertical SOC movements along the soil profile. The $t$ in parentheses means that the corresponding process changes with time.

In this study, we assume steady-state SOC storage for computational efficiency, which is justified by previous research showing that recent disequilibrium effects from climate change and human activities are relatively minor compared to the SOC storage that has developed over thousands of years[31,40]. The steady-state SOC storage $\hat{X}(t)$ can be obtained by letting $dX(t)/dt$ on the left-hand side of equation (3) equal 0. Solving for $\hat{X}(t)$ we obtained:

$$\hat{X}(t) = \left(A\xi(t)K + V(t)\right)^{-1} B(t)I(t) \tag{8}$$

The matrix representation of CLM5 is implemented in PyTorch utilizing vectorized functions to replace most of the for-loops in the original code. Vectorized functions are designed to operate on entire arrays of data simultaneously, rather than processing elements one by one. This enables more efficient computation of SOC predictions in response to changes in parameters. For example, we constructed two vectors for each carbon pool: (1) an environmental scalar vector containing temperature, moisture, oxygen, and depth modifiers that affect decomposition rates, and (2) a decomposition vector containing pool-specific baseline decomposition rates and carbon transfer coefficients. By constructing these vectors for all pools simultaneously (2 vectors for each pool, 7 carbon pools each layer, and 20 layers in total in CLM5), we can directly construct a matrix using vectorized functions. By implementing all mathematical operations (such as addition, matrix multiplication, and matrix inverse) using PyTorch functions, PyTorch can track the gradient of each operation. Using backpropagation, PyTorch can then automatically compute the gradient of the loss function with respect to the learnable weights/biases $\boldsymbol{w}$ of the neural network, differentiating through all the operations in the process-based model (Section 2.1.4).

Equation (7) contains 21 biogeochemical parameters (Table S2) that represent properties of different processes (e.g., transformation and stabilization of SOC, temperature sensitivity of soil respiration, and substrate quality) in the soil carbon cycle [26]. Because those processes are highly variable depending on different climate conditions or soil properties, the values quantifying their properties (i.e., the parameter values) should differ with locations with different environments[26]. Thus, in this study, the neural network embedded in BINN (Section 2.1.1) predicts these biogeochemical parameter values from environmental covariates. The predicted values of the 21 biogeochemical parameters and the environmental forcings (Table S4) are used in Equation (8) to estimate steady-state SOC storage at sites across the Contiguous US.

To mitigate the impact of uncertainties in individual parameters and equifinality on our mechanistic analysis, we aggregated related parameters into functionally distinct model components. Following the method from Tao et al. (2024), these components, including carbon transfer efficiency, carbon input allocation, baseline decomposition, vertical transport rate, environmental modifiers, and plant carbon input, were constructed as weighted averages based on the relative contributions of constituent parameters to their respective carbon fluxes or pools.

### 2.1.3 Loss Function

Because we aim to accurately simulate SOC, our primary loss quantifies the discrepancy between simulated and observed SOC values. Specifically, we use a smooth L1 loss function, which transitions from quadratic behavior near zero to linear behavior beyond a specified threshold β:

$$Smooth\ L1\ Loss(\hat{y}_{profile}, y_{profile}) = \begin{cases} \frac{0.5(\hat{y}_{profile} - y_{profile})^2}{\beta} & if\ |\hat{y}_{profile} - y_{profile}| < \beta \\ |\hat{y}_{profile} - y_{profile}| - 0.5 \times \beta & o.w. \end{cases} \quad (9)$$

where $\hat{y}_{profile}$ represents the simulated SOC profile at all observation depths for a single site by CLM5, $y_{profile}$ denotes the corresponding observed SOC profile at the same site, and $\beta$ is a threshold hyperparameter that determines the transition point between quadratic and linear behaviors of the loss function. The smooth L1 loss function's linear asymptotic behavior makes it more robust to outliers compared to conventional loss functions such as Mean Squared Error (MSE)[41].

We also add an additional hyperbolic cosine loss ($cosh$) term that acts as a regulator, encouraging the neural network to predict biogeochemical parameters within reasonable bounds. Specifically, it penalizes parameter values that deviate substantially from the center of the prior distribution, thereby discouraging biogeochemically implausible extreme values. Eventually, the total loss, $L_{batch}$, is a linear combination of the two losses:

$$L_{batch} = \sum_{profile=1}^{batch\ size} \{Smooth\ L1\ Loss(\hat{y}_{profile}, y_{profile}) + w \sum_{j=1}^{21} cosh\left[\tau(p_j - 0.5)\right]\} \tag{10}$$

where $batch\ size$ is a hyperparameter describing the number of soil profiles processed in each training iteration before performing one backpropagation, $w$ is a weighting hyperparameter that balances the two loss components, $p_j$ represents the predicted biogeochemical parameter from the neural network, and $\tau$ is a scaling factor that controls the strength of regularization by the hyperbolic cosine function. From the hyperparameter grid search, we set $\tau$ to 15 and loss weight $w$ to 100.

While CLM5 simulates SOC dynamics at 20 specific depths, SOC data collected from the field were not necessarily measured at the depth nodes set in CLM5. Thus, in calculating the loss function value, when observations occur at depths between two CLM5 nodes, we employed linear interpolation to estimate simulated SOC values at the observation depths. In cases where observations extend beyond 8 meters (i.e., the deepest node in CLM5 simulations), we used the values at 8 meters as SOC concentration in deeper layers no longer changes much.

### 2.1.4 Backpropagation to Optimize Neural Network Parameters

During training, BINN computes the loss function based on the current predicted SOC and biogeochemical parameters; the loss quantifies how poorly its current predictions match the SOC observations and prior knowledge. Through backpropagation, the loss signals propagate backwards through the entire BINN structural chain: first through the CLM5 matrix equations that generate modeled SOC, then through the biogeochemical parameters, and finally to the neural network that predicts these biogeochemical parameters. At each step, PyTorch uses the chain rule to automatically compute the gradient of the loss function with respect to each learnable NN component (e.g. $w$ in Equation 1 and $\gamma'$ in Equation 2). These gradients indicate how each component can be adjusted to increase or decrease the loss. The NN components are adjusted slightly in the direction that decreases the loss, and then the above process is repeated. The differentiability of the process-based model (CLM5 in this case) enables this continuous gradient flow and thus allows the neural network to learn parameter values that produce better SOC predictions.

### 2.1.5 Monte Carlo dropout

We used Monte Carlo (MC) dropout during BINN training to approximate the model's predictive uncertainty[42]. MC dropout randomly disables a fraction of neurons (with probability equals to the dropout rate) within the neural network during training. This process acts like training an implicit ensemble of subnetworks and improves generalization and robustness of BINN.

When the dropout rate is set larger than 0, the posterior distributions of the predicted biogeochemical parameters can be estimated by MC dropout at each site. To be specific, during forward simulation with the best-trained BINN, we kept dropout active and perform 100 stochastic forward passes with different dropout masks, resulting in an efficient approximation of the uncertainty in the model's parameter predictions.

## 2.2 Recovering Biogeochemical Parameters from Synthetic Data

We evaluated BINN's ability to recover the biogeochemical parameters of CLM5 from synthetic SOC data using a 10-fold cross-validation experiment[43]. Unlike real-world observations that contain measurement uncertainties and potentially unresolved processes, the synthetic SOC data across multiple depths used in this study were generated by running CLM5 with prescribed spatially-varying parameter values (obtained from a previous work with PRODA approach[29]; see Section 2.3.2), providing a controlled environment where true parameter values are known (Fig. S2). This synthetic dataset allowed us to quantitatively assess BINN's parameter recovery accuracy by comparing predicted parameters with the known values used to generate the synthetic data. Parameter recovery was assessed under two configurations: one using a subset of the most sensitive parameters as targets, and another using all biogeochemical parameters as targets.

### 2.2.1 First-Order Sobol Sensitivity Analysis

To decide which biogeochemical parameters to modify in the experiment of recovering the biogeochemical parameters of CLM5 from synthetic SOC, we conducted a sensitivity analysis of CLM5 using first-order Sobol method to identify the biogeochemical parameters that have the greatest influence on simulated SOC values.

The sensitivity analysis was conducted on SOC simulations at various soil-depth ranges, including 0-0.3 m, 0.3-1 m, >1 m, and the entire soil profile (0-8 m). SOC simulations at each layer by CLM5 were aggregated based on the node depths falling into the above-mentioned depth ranges. Specifically, layers 1-6 were used to calculate SOC between 0-0.3 m, layers 7-9 for SOC between 0.3-1 m, and layers 10-20 for SOC greater than 1 m. Simulations from all 20 layers were summed up to calculate SOC across the whole soil profile. The variance and sensitivity for each depth range were calculated based on SOC values derived from the individual layers mentioned above.

For this analysis, we randomly selected 512 sites across the Contiguous US and employed the first-order approximation method. We first determined the unconditional variance $V(SOC)$ from the model output when all the 21 biogeochemical parameters ($p$) in CLM5 were allowed to vary freely within their initial ranges from their prior ranges. Specifically, we randomly sampled the biogeochemical parameter values 1000 times in their initial ranges at each site *s* and depth range *d*, ran the model, and calculated the variance of the simulations, which was considered the unconditional variance $V(SOC_{s,d})$.

Next, we estimated the conditional expectation of the variable $SOC_{s,d}$ for each biogeochemical parameter $p_i$ ($i$ = [0, 20]) at each site *s*. We randomly selected a value (${p_i}^*$) for each biogeochemical parameter $p_i$ from a uniform distribution within its prior range. For the remaining biogeochemical parameters ($p_j$: $j \neq i$), we produced 1000 random settings from uniform distributions within their respective prior ranges. Using the sample of 1000 biogeochemical parameter sets, we estimated the conditional expectation $E\left(SOC_{s,d} \middle| p_i = {p_i}^*\right)$. We repeated this sampling process for 100 randomly selected values of $p_i$ and used the results to estimate the variance $V(E\left(SOC_{s,d} \middle| p_i\right))$. This value quantifies the variance in the output variable $SOC_{s,d}$ as a result of modifying the biogeochemical parameter $p_i$. We discarded the simulations when *NaN* values appeared due to randomly sampled biogeochemical parameter sets. Finally, a first-order Sobol sensitivity index $S_{i,s,d}$ was calculatedfor each biogeochemical parameter $p_i$ at each site *s* and depth *d* as:

$$S_{i,s,d} = \frac{V\left(E\left(SOC_{s,d} | p_i\right)\right)}{V\left(SOC_{s,d}\right)} \tag{11}$$

The final sensitivity value for each biogeochemical parameter was obtained by averaging the sensitivity values across all the randomly selected sites across all depths (Fig. 2) and individual depth ranges (Fig. S3). Error bars represent the mean sensitivity ± one standard deviation across sites.

Our analysis identified the five most sensitive parameters: the parameter "beta" governs the allocation of carbon inputs across soil depths; the parameter "w-scaling" represents the influence of soil water on SOC decomposition; the parameter "tau4s3" represents the decomposability of the passive SOC pool; the parameter "fs1s3" indicates the efficiency of carbon transforming from active SOC to passive SOC; and the parameter "efolding" quantifies the impacts of soil depth on SOC decomposition.

### 2.2.2 Simulation Experiment to Recover Biogeochemical Parameters from Synthetic Data

We selected the five most sensitive parameters identified from the sensitivity test to minimize equifinality, which is a phenomenon that a similar simulation output is generated with different combinations of biogeochemical parameter values[44], for the parameter recovery experiment.

To test the recovery efficiency of the biogeochemical parameters with BINN, we modified the final layer of the neural network by reducing the number of neurons from 21 to 5 to predict these five biogeochemical parameters. Combining the five biogeochemical parameters predicted by BINN and the remaining 16 biogeochemical parameters from the prescribed parameter values, BINN was able to simulate SOC values and update itself through backpropagation. After training BINN with the synthetic SOC dataset, we compared the five parameters predicted by BINN with their prescribed parameter values in the testing dataset to evaluate the accuracy of BINN in retrieving the prescribed biogeochemical parameters. To examine the recovery efficiency beyond the top five parameters, we repeated the experiment with all 21 parameters following the same processes.

### 2.2.3 10-fold cross-validation

To conduct 10-fold cross-validations on the simulations, the entire dataset was randomly divided into ten equal-sized subsets. In each iteration, 8 subsets were used for training, 1 subset for validation, while the remaining subset served as the test set. Specifically, the training subsets were utilized for gradient-based optimization, the validation subset served as an independent check to guide hyperparameter tuning, best model selection, and early stopping to prevent overfitting, and the test subset was reserved for an unbiased assessment of the BINN's generalization capability. This process was repeated ten times, with each subset serving as the test set once. The performance metrics, including Nash–Sutcliffe modelling efficiency coefficient (NSE) and Pearson correlation coefficient (r), were calculated for each iteration. Final performance evaluations were conducted by averaging the metrics across all iterations, and grid-level predictions were averaged across the ten iterations. This cross-validation approach provides a robust assessment of BINN's internal predictive stability by testing its performance on multiple independent datasets, reducing the impact of data partitioning bias and thus enabling evaluation of model stability across different training-testing combinations.

### 2.3 BINN performance with real-world SOC Observations

We evaluated the performance of BINN by comparing BINN's SOC predictions with observed SOC across the contiguous US (a total of 25,925 profiles) (Fig. 1b).

#### 2.3.1 Data Preparation

We processed SOC observations from the World Soil Information Service (WoSIS) following Tao et al. (2020, 2023)[29,31]. Each profile (i.e., each site) may have SOC observations at multiple depths. Only profiles with at least three observations were kept, yielding 25,925 profiles (169,104 SOC measurements) across the contiguous US. We used 60 environmental covariates at each site from Tao et al. (2023)[31] as input to BINN (Table S1). To achieve better training effectiveness, we normalized all the environmental covariates to the interval [0, 1] according to their maximum and minimum values, except the categorical covariates.

We applied eight types of forcing data to drive CLM5 SOC simulations: mean annual net primary productivity (NPP), active soil layer depth from the previous year and current year, the number of soil layers reaching bedrock, the soil oxygen scalar for decomposition, the soil nitrogen scalar for decomposition, soil temperature, and soil water potential. These forcings were derived from 20 years of monthly CLM5 simulations at steady state using a preindustrial forcing (that is, I1850Clm50Bgc) at 0.5° resolution.

The 10-fold cross-validation divided the whole dataset randomly into 10 folds, and we took one-fold (i.e., 10%) data as the testing dataset in each iteration. The remaining data were further split into training (8/9) and validation (1/9) sets.

### 2.3.2 PRODA

We used PRODA as a baseline to compare SOC simulation skill and derived process components with BINN. PRODA couples site-level Bayesian calibration of CLM5 with a neural network that maps environmental covariates to the calibrated parameters.

The site-level data assimilation was conducted by an adaptive Metropolis Markov Chain Monte Carlo (MCMC) algorithm with 20,000 iterations for the test run and 50,000 iterations for the formal run. The posterior distribution for each parameter was generated during the second half of the formal run. The mean values of the posterior distributions of the parameters were calculated as the final point estimates.

To generalize these site-level estimates to the contiguous US, we trained a fully-connected multilayer neural network using the same 60 environmental variables as predictors. The neural network used in the final training consisted of four hidden layers. The node numbers for each hidden layer were 256, 512, 512 and 256, respectively, with *ReLU* as the activation function and *adadelta* as the optimizer. PRODA was trained with 6000 epochs, utilizing an early stopping patience of 1200 epochs, and validated with a 10-fold cross-validation. When comparing with BINN, the test dataset for each fold was set the same as the corresponding fold during BINN training.

To facilitate a direct comparison of model performance between BINN and PRODA (Fig. 4 and 6), we used the identical prior ranges of parameters governing substrate decomposability as used in PRODA. Specifically, the prior ranges were for tau4s1 were [0.0001, 1]), tau4s2 [1, 50]), and tau4s3 ([200, 1000]). Aside from this specific comparative analysis, the values of prior parameter ranges in Table S2 were used for all other analyses.

The robustness of the BINN framework to the variations in prior range selection was further evaluated by adjusting the prior ranges by 10 percent where the flux related parameters remained within their physical bounds. The results in Table S3 indicate that BINN is robust to the test of prior ranges as evidenced by the consistent NSE across different settings of modeling choices.

### 2.3.3 Summary Statistics

We calculated the Nash–Sutcliffe modelling efficiency coefficient, NSE, of simulated SOC to evaluate the effectiveness of SOC predictions by BINN following the equation:

$$NSE = 1 - \frac{\sum(obs_i - simu_i)^2}{\sum(obs_i - \overline{obs_i})^2} \tag{12}$$

where $obs_i$ is one of the SOC observations, $\overline{obs_i}$ is the mean of the SOC observations, and $simu_i$ is the simulated SOC by CLM5 embedded in BINN.

We used the Pearson correlation coefficient, *r*, between the predicted and prescribed biogeochemical parameters to evaluate the effectiveness of BINN in recovering each of the five biogeochemical parameters as follows:

$$r = \frac{\sum[(para_{BINN} - \overline{para_{BINN}}) \times (para_{true} - \overline{para_{true}})]}{\sqrt{\sum(para_{BINN} - \overline{para_{BINN}})^2 \times \sum(para_{true} - \overline{para_{true}})^2}} \tag{13}$$

where $para_{BINN}$ is the biogeochemical parameter predicted by BINN, $para_{true}$ is the biogeochemical parameter previously prescribed at the same site, $\overline{para_{BINN}}$ is the mean of this biogeochemical parameter predicted by BINN, and $\overline{para_{true}}$ is the mean of this prescribed biogeochemical parameter.

### 2.3.4 Normalized Difference

For each site, we averaged the simulated SOC across the 10 cross-validation folds and subtracted the observed SOC to obtain a mean difference. To improve interpretability over space, we applied a signed percentile normalization: positive differences were scaled to [0,1] by dividing by the 99th percentile of positive deviations, and negative differences were scaled to [−1,0] by dividing by the 99th percentile of negative deviations. This approach preserves the sign and relative magnitude of biases while preventing a small number of extreme values from dominating the colored scale.

### 2.4 Computational Efficiency

We compared computational efficiency between BINN and PRODA on the same dataset (2,000 soil profiles) and the same personal computer, reporting their wall-clock time and peak resident set size (RSS). For BINN, we evaluated two implementations and recorded their respective runtime and peak RSS: (i) a matrix version using the original CLM5 formulation and (ii) a vectorized version that removes explicit for-loops. Each implementation was trained for 300 epochs (no early stopping) to standardize timing and ensure convergence on the training set. For PRODA, the computational bottleneck is the site-level MCMC calibration of CLM5 parameters. We measured MCMC runtime at 10 randomly selected sites (with 20,000 test iterations and 50,000 formal iterations per site) and extrapolated to 2,000 sites. Because per-site memory usage is stable across sites, peak RSS was taken from these 10-site runs. PRODA's second stage using a neural network that maps environmental covariates to the MCMC-optimized parameters was trained for 6,000 epochs, with an early stopping patience of 1200 epochs. We recorded its peak RSS and reported PRODA's overall peak RSS as the maximum of the MCMC stage or the NN stage.

## 3. Results

### 3.1 Parameter Recovery Accuracy

When BINN predicted the five most sensitive biogeochemical parameters, the recovered parameter values exhibited high consistency with the prescribed biogeochemical parameters used during synthetic data generation, achieving an average correlation coefficient of 0.88 across the 10-fold cross-validations (Fig. 3g). Moreover, BINN achieved an average NSE of 0.99 on the test dataset when comparing simulated SOC with synthetic SOC (Fig. 3g). In the cross-validation iteration with the median NSE of simulated SOC, the BINN-recovered parameter "efolding", representing the depth scalar, had a correlation coefficient of 0.92 in comparison with the prescribed parameter values (Fig. 3a). The parameter "tau4s3", representing the baseline turnover time of the passive SOC pool, showed a correlation coefficient of 0.93 (Fig. 3b). The "fs1s3" parameter, indicating the transfer fraction from fast to passive SOC pools, achieved a correlation coefficient of 0.82 (Fig. 3c). The "w-scaling" parameter, which is the scaling factor of soil water, had a correlation coefficient of 0.9 (Fig. 3d). Lastly, the "beta" parameter that governs the vertical distribution of carbon inputs exhibited a correlation coefficient of 0.83 (Fig. 3e).

Extending recovery to all 21 parameters revealed heterogeneous identifiability that is consistent with their influence on SOC storage (Fig. S4). Parameters with low sensitivity (e.g., cryo, taucwd) showed weak correlations between retrieved and prescribed values and were typically distributed around the midpoints of their prior ranges, whereas high-influence terms (e.g., w-scaling, fs1s3) remained well recovered.

In addition, to better address the equifinality issue in BINN, we implemented MC dropout[42] to approximate the posterior distributions of parameters at each site (Fig. S5). Although point estimates of parameters may not exactly match the prescribed values in parameter recovery experiment, the posterior distributions of parameter values by BINN frequently cover the prescribed values. We further quantified the site-level coverage (i.e., the fraction of prescribed values falling into the BINN-predicted posterior distributions) in Fig. S6, which indicated the issue of equifinality is effectively identified and accounted for within the BINN framework, as the uncertainty intervals consistently encompass the ground-truth parameter values at most sites.

### 3.2 Performance in Predicting Real-world SOC Observations

After BINN optimization, we used its predicted biogeochemical parameters to calculate six model components representing different processes in soil carbon cycle over the contiguous US: carbon transfer efficiency, baseline decomposition, environmental modifier, carbon input allocation, vertical transport rate, and plant carbon inputs. The comparisons between BINN-retrieved six model components and those generated from PRODA, along with their spatial patterns, are shown in Fig. 4. Note that the plant carbon input component was identical between BINN and PRODA (Fig. 4p-r), due to the use of the same NPP forcing data. The spatial distributions of the other five components were similar between BINN and PRODA, with an average correlation coefficient of 0.86.

The carbon transfer efficiency predicted by BINN, which quantifies the weighted average ratio of decomposed carbon being transferred from one carbon pool to another relative to the total carbon decomposition, displayed similar spatial patterns as PRODA's predictions (Fig. 4a–b), with a correlation coefficient of 0.92 (Fig. 4c). Baseline decomposition, which describes the substrate decomposability of each soil pool, showed similar spatial patterns across the contiguous US for both BINN and PRODA as well (Fig. 4d–e), with the correlation coefficient of 0.91 (Fig. 4f).

The environmental modifier predicted by BINN achieved a correlation coefficient of 0.82 with PRODA's results (Fig. 4i). Their spatial patterns were nearly identical across the contiguous US, with lower values in the northwestern region and values gradually increasing toward the southeastern region (Fig. 4g–h). While both methods simulated higher carbon input allocation in the central US, BINN predicted slightly higher values in the western and central parts of the contiguous US compared to PRODA (Fig. 4j–k), resulting in a relatively high average value. Even so, the correlation coefficient between the two approaches still reached 0.71 (Fig. 4l). BINN and PRODA also predicted vertical transport rate with nearly identical spatial distributions (Fig. 4m–n), with a high correlation coefficient of 0.92 (Fig. 4o).

When applying the predicted parameters to simulate SOC, BINN showed an average NSE of 0.71 in predicting the observed SOC across the 10-fold cross validations (Fig. 5c), with an average NSE of about 0.57 on the test datasets (Fig. 5d). The training and validation NSE values recorded throughout model training at each epoch indicated a validation NSE of 0.63 (Fig. S7). The map of the normalized difference (Fig. 5a) identifies geographic regions where BINN's predictive performance is limited, showing systematic underestimation in the North-Central region and overestimation within the Rocky Mountains., which persisted across the 10-fold cross-validation ensemble (Fig. 5b).

Overall, BINN demonstrated in-domain predictive accuracy comparable to PRODA in predicting SOC across the contiguous US (Fig. 6a). However, PRODA showed a tendency toward negative bias (underestimation) over broad areas, whereas BINN's normalized difference was closer to zero (Fig. 6b). In the north-central and eastern US, where PRODA mainly underestimated observed SOC, BINN exhibited fewer geographical biases (Fig. 6c–d).

### 3.3 Computational Efficiency

On a single CPU, BINN with for-loops required 52.5 hours (Fig. 7) to complete 2,000 profiles, whereas the vectorized version took only 10.5 hours (5× faster). Moreover, MCMC alone took 574.7 hours, leading to a total of 577.7 hours for PRODA when including ~3 hours of neural network training. Thus, BINN is approximately 57 times faster than PRODA and can be further accelerated via PyTorch DDP, which enables parallel training across multiple CPUs. While both approaches achieve parameter interpretability (Section 3.2), BINN does so more efficiently by directly integrating the process-based model into the neural network architecture, eliminating the need for computationally intensive site-by-site MCMC optimization. In addition, BINN with vectorized CLM5 yields the lowest peak memory usage among all the methods (Fig. S8).

## 4. Discussion

Our study demonstrates that BINN is a novel approach to retrieve spatially varying model parameters from big data and predict their spatial distributions for determining SOC dynamics over the contiguous US and thereby facilitate process understanding. By embedding a matrix form of process-based model into BINN, BINN can significantly enhance computational efficiency during parameter optimization compared to conventional data assimilation techniques. Additionally, the incorporation of MC dropout enables the estimation of posterior distribution for each parameter, facilitating rigorous uncertainty quantification within this hybrid modeling framework.

### 4.1. BINN's ability to retrieve and predict biogeochemical parameters

In this study, BINN's ability of retrieving biogeochemical parameters was first validated through a parameter recovery experiment, which used synthetic SOC data generated by CLM5 with prescribed parameter values across the contiguous US. To minimize the effects of equifinality, we conducted a sensitivity analysis using a first-order Sobol index and focused on the five most sensitive biogeochemical parameters, which are also those typically well constrained by Bayesian calibration. It is important to note that while the first-order Sobol index identifies the direct contribution of individual parameters to model variance, it may not fully account for all nonlinear interactions within the system. Therefore, future research involving higher-order sensitivity analysis would be beneficial to further explore complex parameter relationships.
As expected, BINN recovered these target parameters with high correlations with their respective prescribed values. Moreover, BINN reproduced the synthetic SOC patterns well using only environmental covariates and synthetic SOC, without using a Bayesian method. These results suggest that BINN learns causal relationships between environmental covariates and SOC simulations consistent with the underlying dynamics, rather than simply exploiting spurious correlations[26].

We then extended recovery to all 21 parameters; the results showed that highly sensitive parameters remained well recovered. Although parameters with low influence on SOC showed weak recovery by BINN, this is a common limitation also observed in Bayesian approaches when data are weakly informative. To overcome this limitation, future applications could integrate complementary data streams beyond SOC, which might provide additional constraints necessary to inform those less-sensitive parameters. Alternatively, the inherent uncertainty of high-dimensional parameter spaces can be mitigated by aggregating individual parameters into functionally related model components. As demonstrated in this study, grouping parameters with similar process-based meanings and weighing them by their relative contributions to total SOC help characterize the underlying biogeochemical processes more powerfully (Fig. 4). This approach effectively prioritizes the most influential parameters; as weakly recovered parameters generally exert minimal impact on total SOC storage. Such parameters are assigned lower weights within the calculated model components, thereby mitigating the impact of poorly constrained parameters on our mechanistic interpretations.

To further characterize and mitigate equifinality, we applied MC dropout to approximate the posterior distributions of site-level parameters. Although point estimates did not necessarily match the prescribed values exactly, the posterior distributions frequently cover the ground truth (i.e., the prescribed parameters in this case). This demonstrates BINN's ability to identify and account for equifinality issues. By generating a distribution of potential parameter values instead of a single point estimate, the framework provides a robust quantification of uncertainty that carries through from the biogeochemical parameters to the final model simulations.

We further tested BINN with real-world SOC observations across the contiguous US and quantified the six model components of CLM5. BINN-retrieved model components, calculated from the estimated biogeochemical parameters, showed good agreement with those generated by PRODA. Since Bayesian optimization as used in PRODA is widely accepted in earth system modeling for parameter optimization through data assimilation, the agreement between BINN and PRODA suggests that BINN can effectively capture spatial variations in key model components while maintaining physical interpretability. This capability enables BINN to evaluate the relative importance of different processes controlling SOC storage within the model and data domain, as done with PRODA, while offering computational advantages through its integrated neural network architecture.

While both models show comparable NSE values for SOC simulations (Fig. 6a), their spatial error structures differ significantly. PRODA demonstrates a widespread tendency to underestimate SOC stocks, whereas BINN's normalized differences are centered nearer to zero, indicating reduced systematic bias (Fig. 6b). This difference likely stems from the architectural distinction between the two frameworks: while the neural network in PRODA is trained to replicate site-level parameter distributions derived from a separate site-level data assimilation step, BINN's end-to-end framework directly optimizes the neural network to minimize the deficit in SOC simulations across the US continent. By coupling the parameter inference and the physical simulation within a single objective function, BINN can more effectively mitigate spatial biases that might otherwise arise from the multi-step training process used in PRODA.

In addition, all model components predicted by BINN show high correlations with those predicted by PRODA. However, the estimated carbon transfer efficiency in BINN is slightly lower than in PRODA, primarily due to higher carbon input allocation to deeper soil layers in BINN. Since shallow layers typically maintain larger carbon stocks than deeper layers, the weighted carbon transfer efficiency in BINN is lower to compensate for the reduced input allocation in those shallow layers. This minor bias can also be attributed to the distinct architectural designs between BINN and PRODA and the resulting differences in optimized parameters.

However, even though BINN shows less underestimation than PRODA, spatial bias still persist within the North-Central and Rocky Mountain domains (Fig. 5a). These biases likely stem from a fundamental mismatch between the spatial resolution of the environmental covariates that derived from global gridded products (Table S1), and the local-scale drivers of each SOC observation. In the North-Central region, the high spatial density of SOC observations reveals sub-grid variability that coarse gridded covariates cannot adequately resolve. Conversely, the overestimation observed in the Rocky Mountains highlights a scale-dependent limitation where gridded inputs fail to represent the complex topographic heterogeneity inherent to mountainous terrain. Consequently, while the BINN framework effectively captures broad environmental gradients, its localized precision remains constrained by the granularity of the available geospatial drivers.

### 4.2. BINN's Computational Efficiency

BINN shows significant improvement in computational efficiency compared to PRODA while performing similar functionality in terms of retrieving parameters and predicting the spatial distributions of SOC storage and its components from big data. Compared to PRODA, BINN reduces computational time by more than 50-fold in a test with 2,000 profiles, while keeping memory usage the lowest. PRODA requires running a Bayesian optimization algorithm for each site independently; it does not use gradients to optimize parameter values. Instead, it only perturbs parameters randomly and checks whether the accuracy has been improved or not.

By contrast, BINN reimplements the process-based model, CLM5, in a differentiable way using PyTorch, and leverages this differentiability to rapidly find the optimal parameters (for all sites simultaneously) that accurately simulate SOC observations (Table S5). Additionally, we used vectorized functions in PyTorch to replace for-loops in the original matrix form of CLM5 model, which further enhances computational efficiency. BINN can also utilize PyTorch DDP to parallelize computations, allowing much faster iterations and thus greatly saving real physical time. Meanwhile, the memory usage of BINN can be kept bounded via smaller batches and gradient checkpointing, whereas site-by-site Bayesian samplers must maintain full state per chain, which tends to increase peak RSS despite longer running time. High computational efficiency is also more environmentally friendly, saving energy when dealing with large datasets.

### 4.3. BINN's Facilitation of Mechanistic Understanding

A process-based model is an abstraction of a real-world system and represents the processes that govern the system's dynamics, yet such model-based predictions often fit with empirical observations poorly[45]. This discrepancy arises because complex systems, such as the terrestrial ecosystems, contain numerous mechanisms regulating carbon cycling. Although some of these mechanisms are well-understood, many remain unresolved. Process-based models may explicitly represent those well-understood processes in their structure while these processes need to be specified via parameterization at given locations under given conditions to represent the heterogeneity of environmental conditions[26]. Without taking advantage of the extensive information present in observations, these model parameters are usually not well constrained.

BINN learns site-specific parameters from environmental covariates while the process-based model executes the forward simulation. This end-to-end setup improves model fit to observations by yielding spatially varying parameters. When these parameters are properly constrained by empirical data, they can more accurately reflect the unresolved biogeochemical processes, enabling models to better simulate ecosystem behavior. After training, we further transform the learned parameters into model components (as in Fig. 4), thereby linking learned parameters to individual processes. The model components from BINN reveal interpretable patterns (e.g., baseline decomposition and carbon transfer efficiency) consistent with prior work using Bayesian-based methods[31].

However, it is important to recognize that the mechanism understanding inferred through this framework is inherently tied to the CLM5 model structure and the BINN-optimized parameters. While these optimized parameters provide a mechanically consistent representation of different processes within the process-based model, the resulting mechanistic discovery may still require further independent validations, such as observational data from satellites, sensors, and field and laboratory experiments. Nonetheless, BINN-retrieved mechanism understanding is much better grounded in observations than these gained by the process-based models or data alone.

BINN's potential for advancing scientific understanding could be extended well beyond SOC dynamics to various fields of research in biogeochemistry and ecology. Whenever current scientific understanding of a biogeochemical system can be mathematically formulated in a process-based model, BINN could help uncover mechanisms from big data when relevant observational data are presented. This framework would be particularly valuable for studying complex biogeochemical cycles, such as nutrient cycles, where some processes are well-understood and explicitly represented in models, while others remain unresolved. By combining process-based representations of known mechanisms with large-scale data, BINN could help uncover unknown mechanisms governing biogeochemical cycling, providing a foundation for future empirical studies to validate these model-derived insights.

Furthermore, BINN's flexible architecture might allow integration of diverse data sources. This capability would be particularly valuable for incorporating limited but important datasets, such as isotope measurements, which could reveal spatial and temporal mechanisms in soil carbon dynamics despite their scarcity. Even if these measurements are only available at a few sites, BINN could still learn from them by incorporating them in the loss function where they are available. By leveraging multiple data sources, BINN would maximize the potential to facilitate our scientific understanding while maintaining biogeochemical consistency.

**(a)**

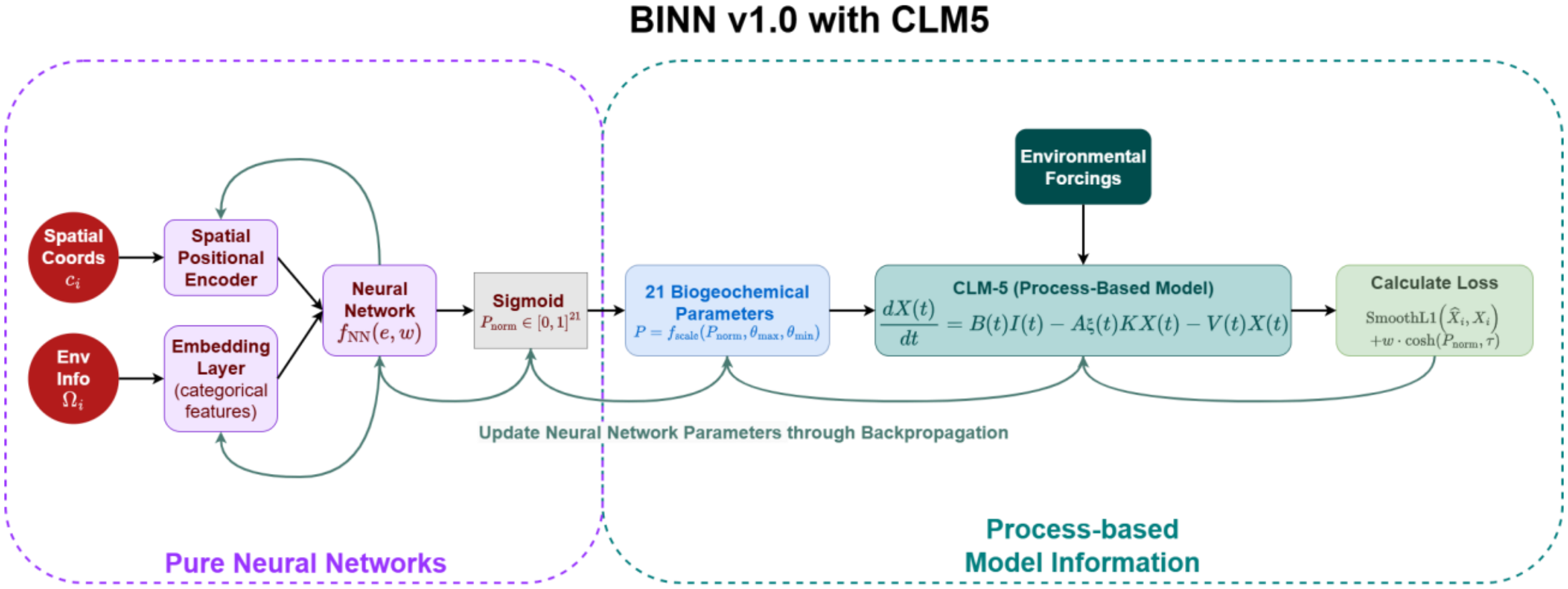

**(b)**

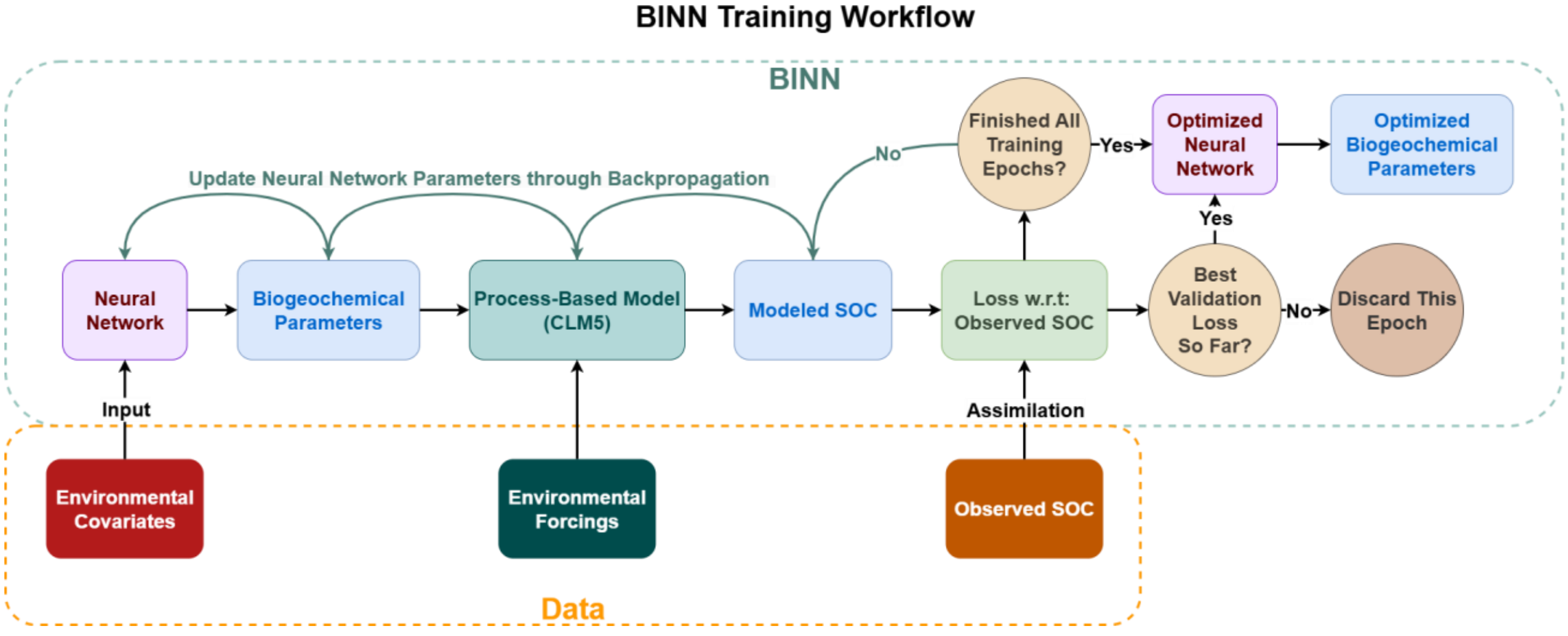

**Figure 1. Schematic diagram of BINN architecture and training workflow.** (a) Detailed BINN structure showing the integration of neural networks with CLM5. A neural network (purple) processes spatial coordinates (red) through a positional encoder and categorical

environmental covariates through an embedding layer. The network outputs are transformed via a sigmoid activation function to generate 21 normalized parameters (grey). These parameters are then scaled by their corresponding prior range (blue) and pass into CLM5 along with environmental forcings to simulate SOC (teal). The model's performance is evaluated using a smooth L1 loss function and a soft prior penalty on biogeochemical parameters (green). The entire framework is differentiable, enabling end-to-end training through backpropagation (teal arrow). (b) Overview of BINN training workflow. Environmental covariates at each site serve as input to the neural network to predict biogeochemical parameters. These parameters, along with environmental forcings, drive the process-based model (CLM5) to simulate SOC. The difference between modeled and observed SOC is used to compute the loss function, which guides neural network parameter updates through backpropagation (teal arrow). This training process continues until reaching the maximum number of epochs or achieving optimal validation performance.

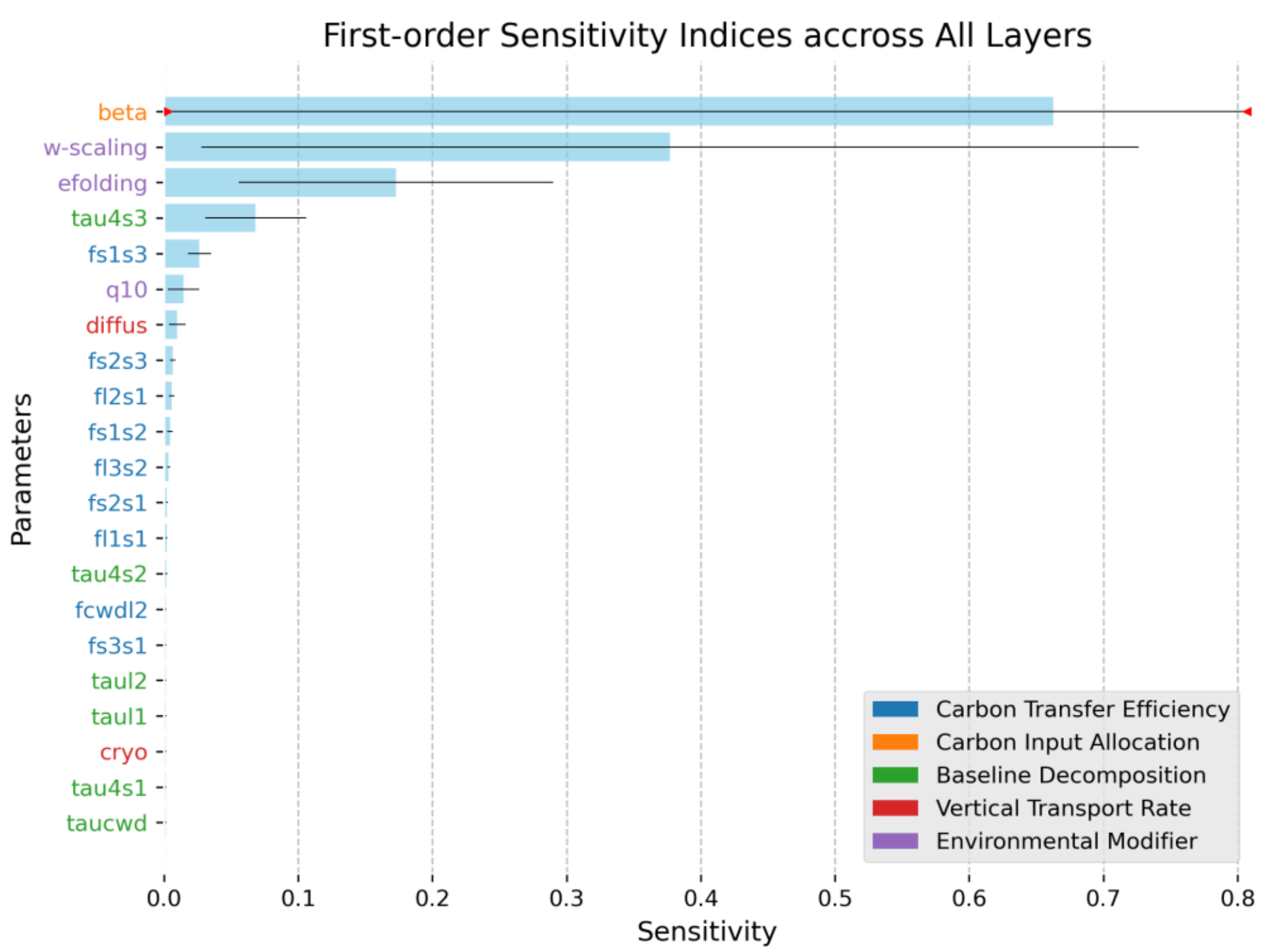

**Figure 2: First-order sensitivity indices of biogeochemical parameters aggregated across all soil layers.** Bars show mean sensitivity, with horizontal error bars representing ±1 standard deviation. Red triangles indicate error bars clipped by the x-axis (uncertainty extends beyond the displayed range). Parameters are ranked (y-axis) by decreasing sensitivity and color-coded by their associated process component (e.g., $K$, $A$, $B$, $\xi$, $V$ in Equation 7). The x-axis represents sensitivity scores, indicating how changes in each parameter influence the model's performance.

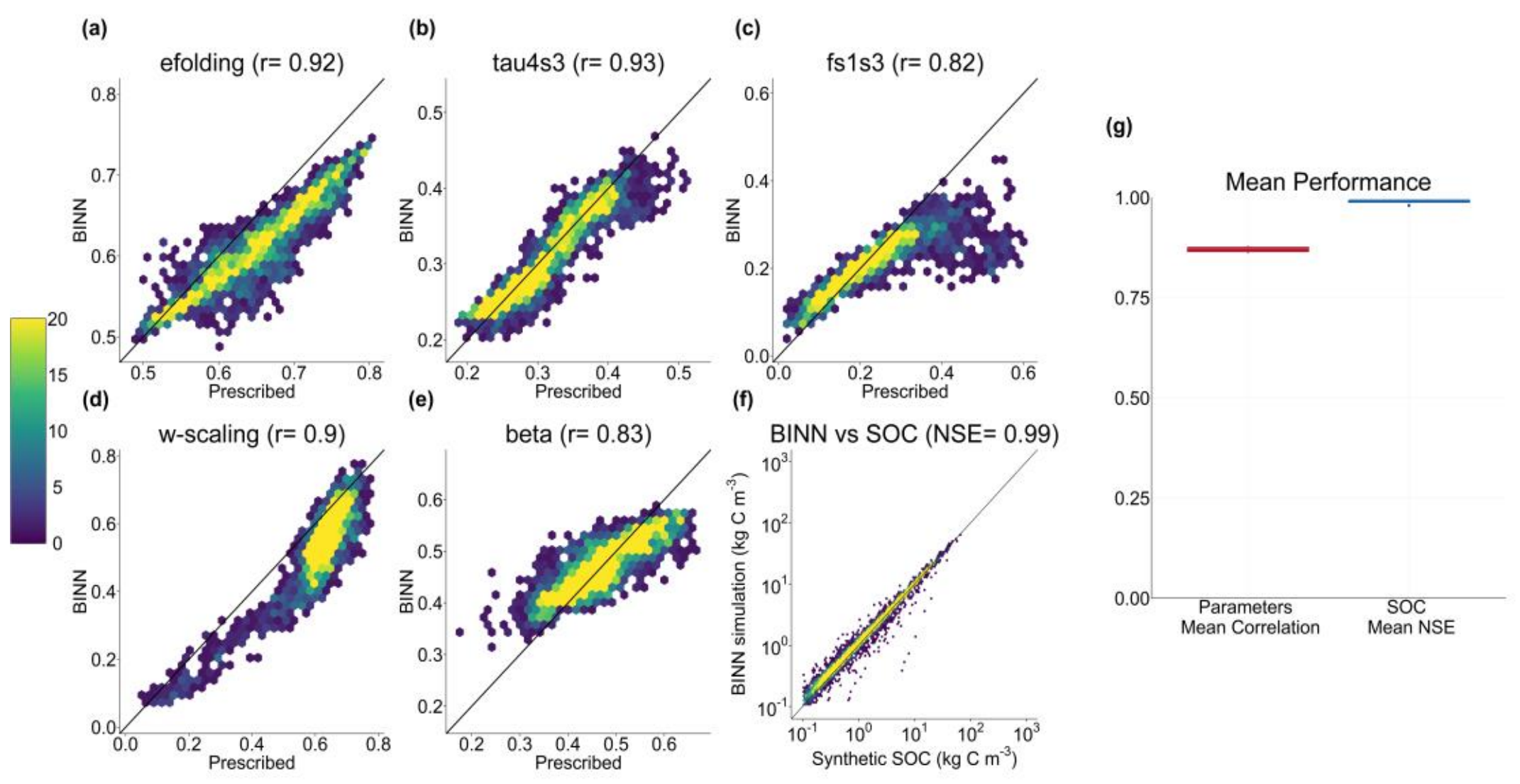

**Figure 3: BINN's performance in the parameter recovery experiment.** Scatter plots comparing the parameter values (a) "efolding", (b) "tau4s3", (c) "fs1s3", (d) "w-scaling", and (e) "beta" predicted by BINN against the prescribed parameter values. The color of each point in the scatter plot represents the number of data points within each hexagonal bin. The correlation coefficient between the predicted and prescribed parameter values is shown in the title of each plot. (f) Comparison of the BINN-simulated and synthetic SOC values, with colors representing

the density of points. (g) Mean performance of BINN in retrieving the five parameters across 10-fold cross validations, as measured by the correlation coefficient between the predicted and prescribed parameter values as well as NSE of the simulated SOC values compared to the synthetic SOC data.

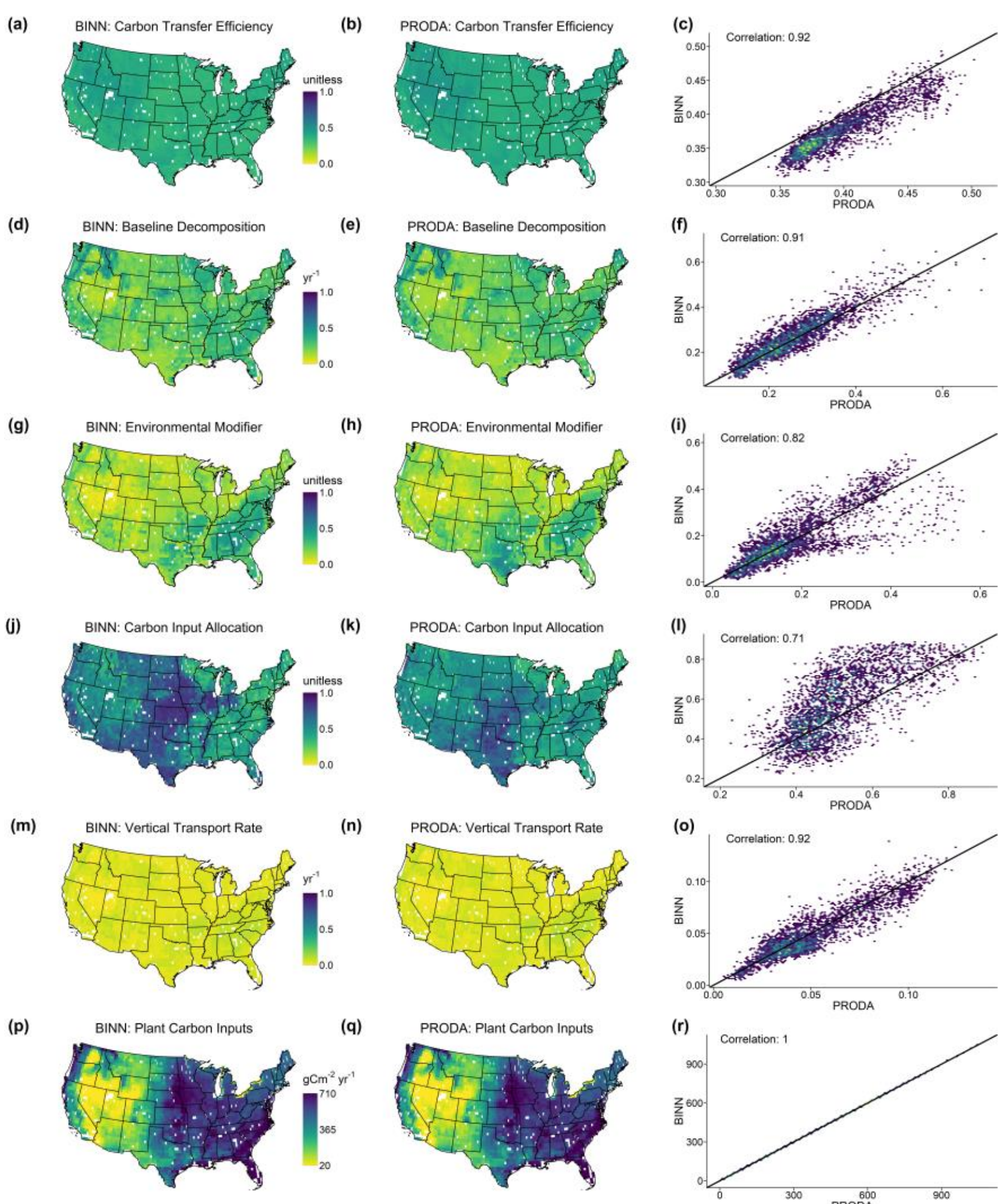

**Figure 4: Comparison of the spatial patterns of model components retrieved by BINN and PRODA.** Maps show BINN (left column panels) and PRODA (middle column panels) estimates for six model components: carbon transfer efficiency (a–c), baseline decomposition (d–f), environmental modifier (g –i), carbon input allocation (j–l), vertical transport rate (m–o), and plant carbon inputs (p–r). Right-column panels (c, f, i, l, o, r) compare model components between BINN (y-axis) and PRODA (x-axis) with 1:1 lines; panel insets report Pearson correlation coefficients. Color scales are harmonized across BINN and PRODA; otherwise, panel-specific scales are noted in the legend.

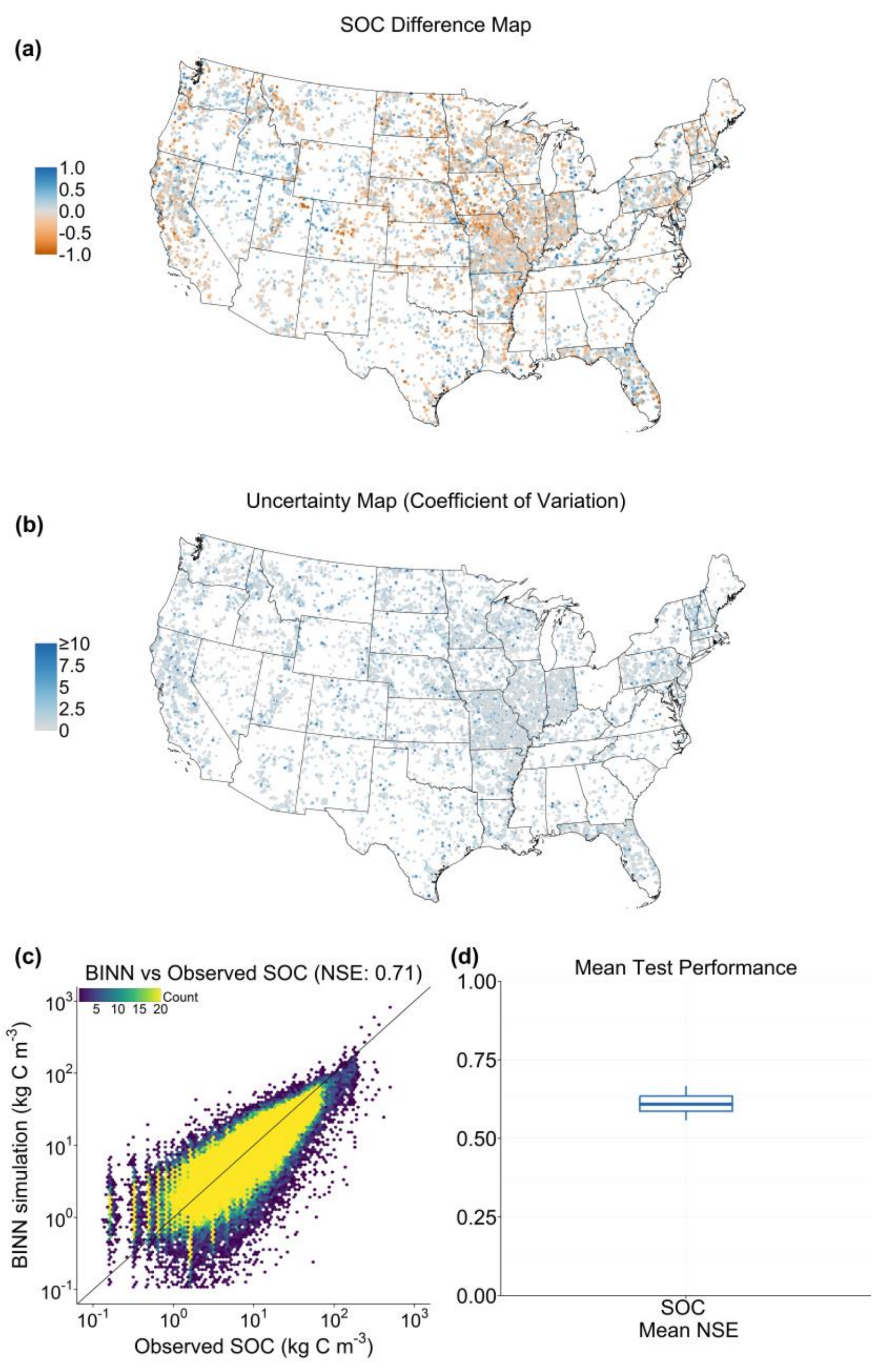

**Figure 5: Comparison of simulated and observed SOC storage using BINN across 10-fold**

**cross-validation training.** (a) Map of normalized difference across all 10 folds. (b) Coefficient of variation of difference across 10 folds, indicating spatial uncertainty. (c) Observed vs. fold-average simulated SOC across all depths and sites. (d) Box plot of test NSE across 10 folds.

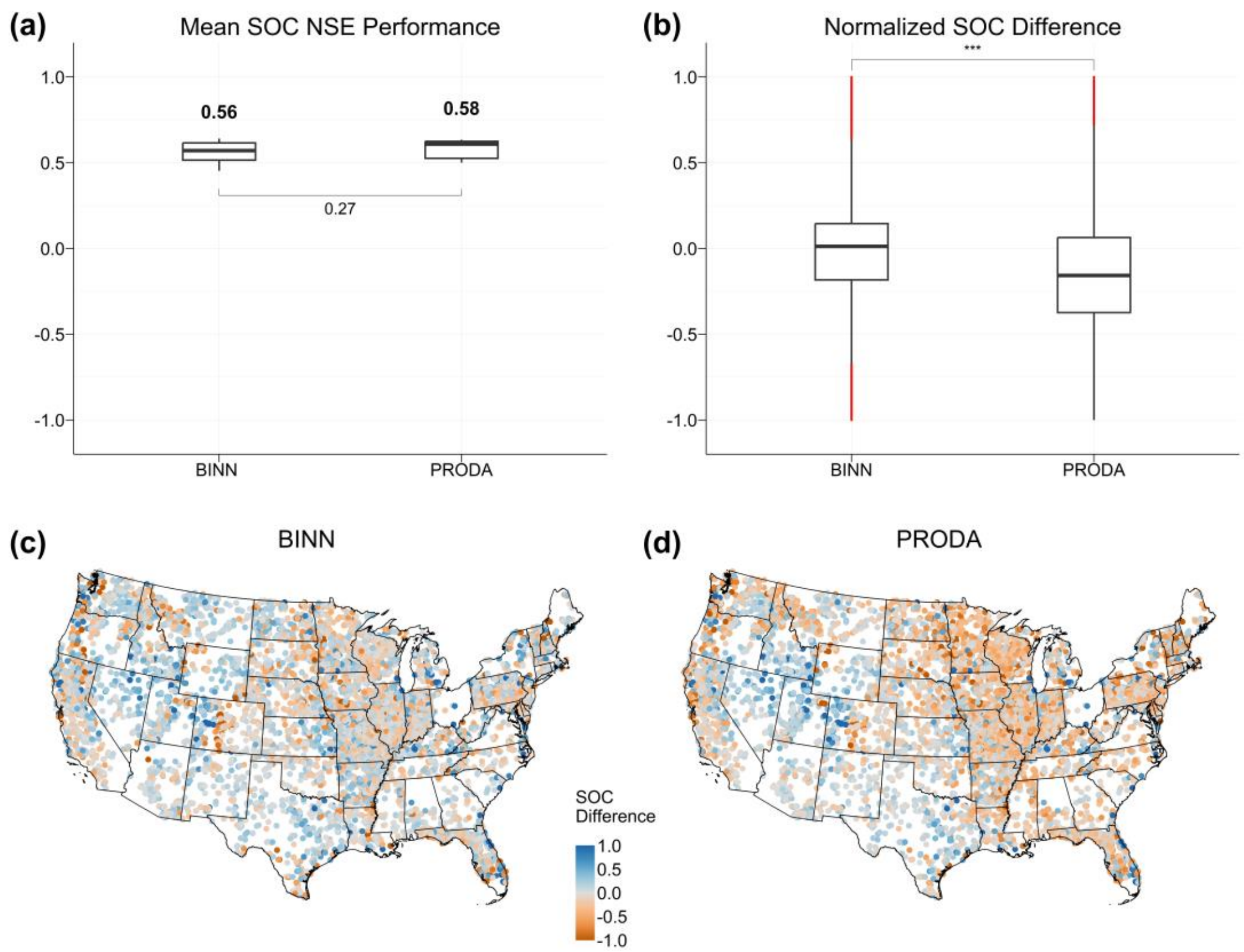

**Figure 6. Side-by-side evaluation of BINN and PRODA across identical 10-fold test sets against the observed SOC.** (a) Mean test NSE per fold for BINN and PRODA. NSE values are shown above the bars; the P value below the bars donates the difference in test NSE between the two methods is insignificant. (b) Mean test-set differences in simulated SOC between the two methods after normalization. The red dots are outliers, and the horizonal significant bar shows that the mean differences in SOC between two methods are significantly different. (c–d) Spatial maps of the normalized differences for (c) BINN and (d) PRODA: values near 0 indicate small bias; positive and negative values indicate overestimation and underestimation, respectively.

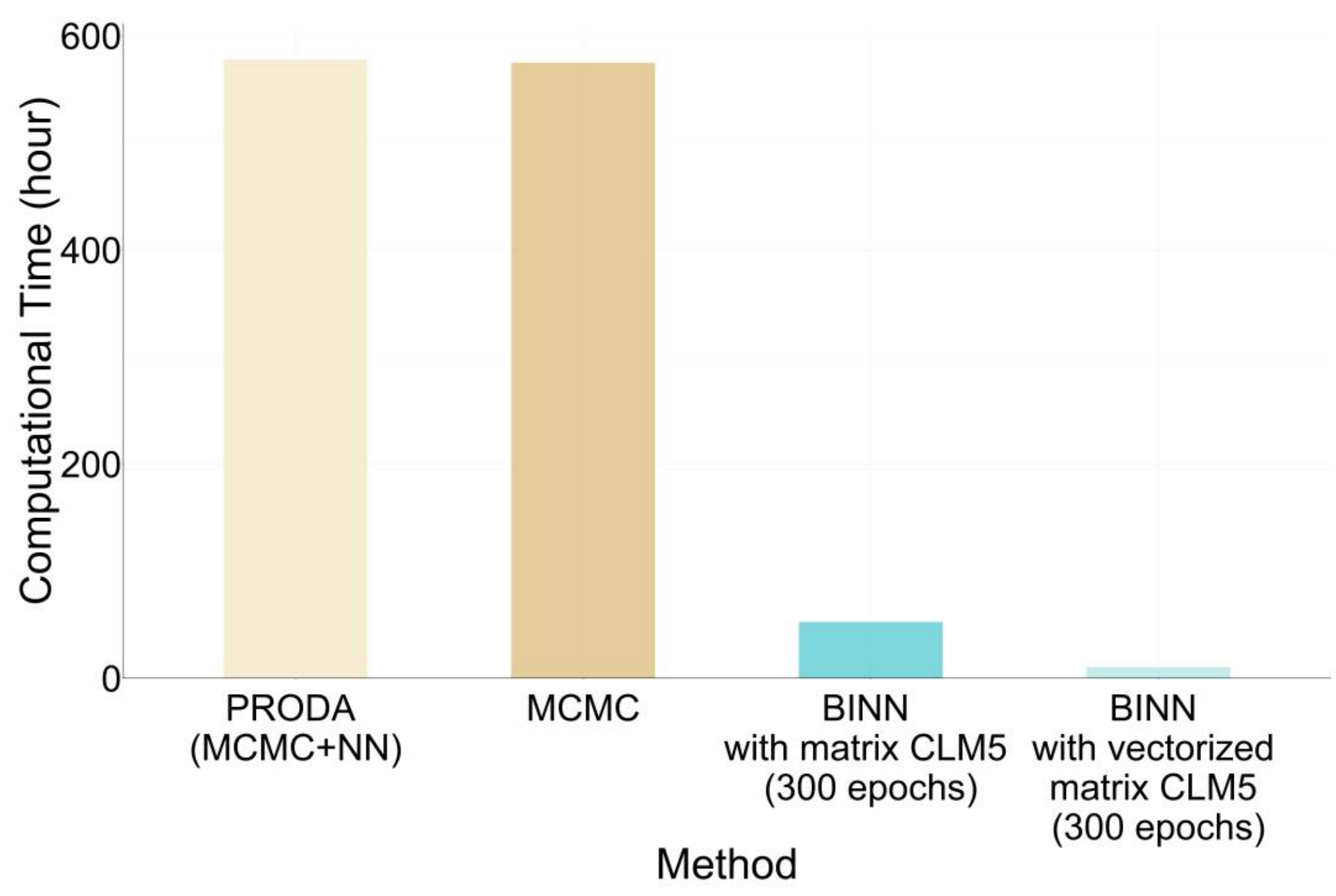

**Figure 7: Computational time to process 2,000 soil profiles with CLM5.** Wall-clock runtime (hours) for PRODA (MCMC + NN), MCMC only, and for BINN using the original matrix CLM5 and a vectorized CLM5 implementation. All runs were executed on the same platform.

## Acknowledgements

This research is supported by AI-LEAF: "AI Institute for Land, Economy, Agriculture & Forestry", funded by the USDA National Institute of Food and Agriculture (NIFA) and the NSF National AI Research Institutes Competitive Award (No. 2023-67021-39829). This research is also partially supported by Schmidt Sciences programs, an AI2050 Senior Fellowship and an Eric and Wendy Schmidt AI in Science Postdoctoral Fellowship; NSF grants (DEB 2242034, DEB 2406930, and DEB 2425290); the US Department of Energy's Terrestrial Ecosystem Sciences Grant DE-SC0023514, subcontract CW55561 from Oak Ridge National Laboratory to Cornell University; the CALS Moonshot Seed Grant program; the "NYS Connects: Advancing Markets for Producers" project, funded by the USDA and the New York State Department of Environmental Conservation; an NSF Research Traineeship (NRT) fellowship in Digital Plant Science (DGE-1922551), and the Air Force Office of Scientific Research (AFOSR, grants FA9550-23-1-0322, FA9550-23-1-0569, FA9550-21-1-0316).

## Data and Code Availability

According to Xu et al. (2026) [46], the code and data for BINN are available at:

https://doi.org/10.5281/zenodo.19237379.

## Author Contribution

HX, YL, FT, and JF: Conceived this study; HX and JF: Developed the methods and performed the analyses; HX: Prepared the first draft of the manuscript; HX, JF, FT, LJ, FY, BH, YS, CG, and YL: Revised the manuscript.

## Conflicts of Interest

The authors declare no competing interests.

## Supplementary Information

### Supplementary Tables and Figures

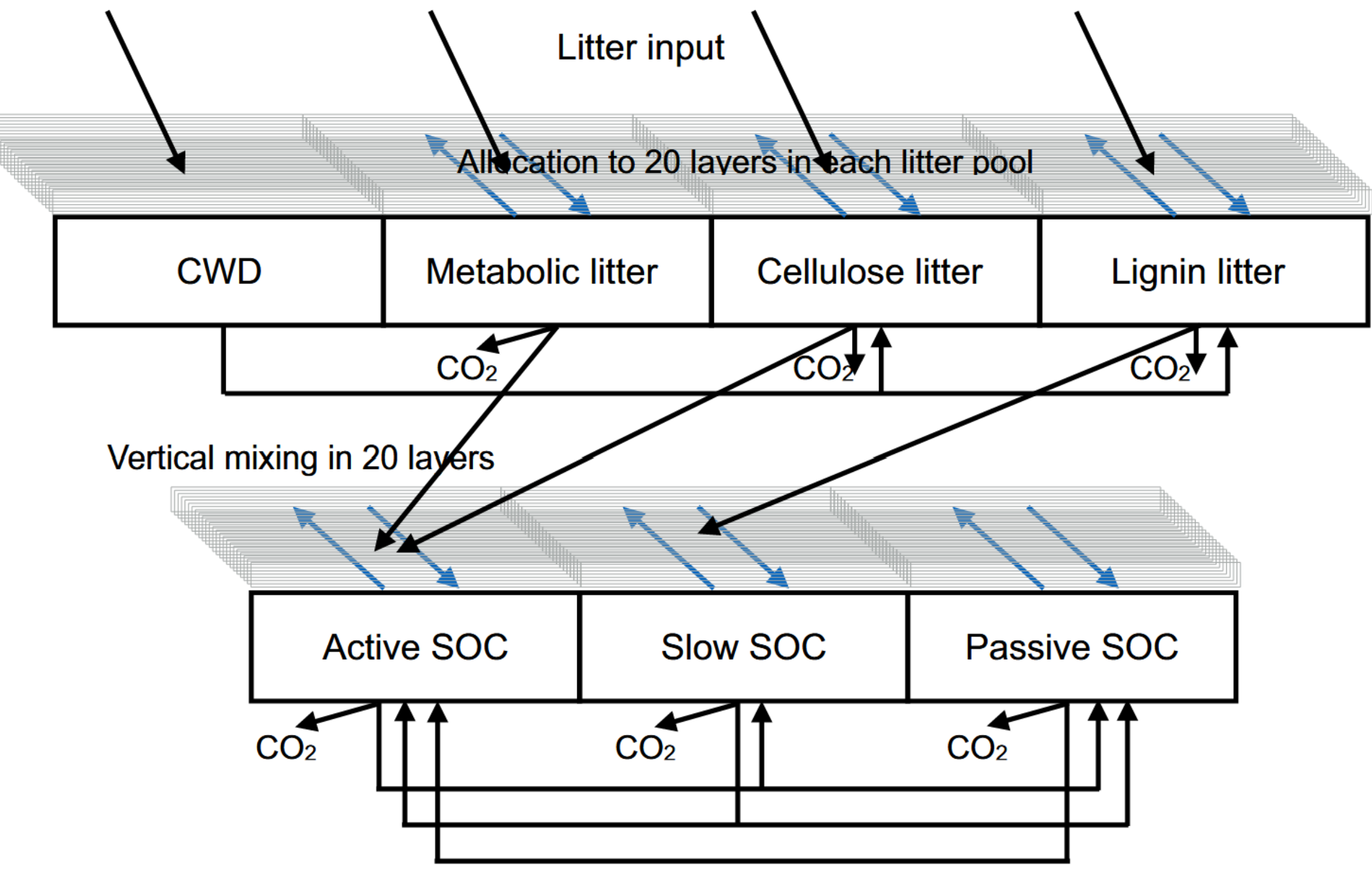

**Supplementary Figure 1:** Model structures of CLM5[1].

**Supplementary Table 1:** Environmental Covariate Data as BINN input

| No. | Variable Name | Data Source | Category | Description |
|---|---|---|---|---|
| 1 | Lon | WoSIS | Geography | Longitude |
| 2 | Lat | | | Latitude |
| 3 | Elevation | NOAA | | Elevation |
| 4 | Abs_Depth_to_Bedrock | (Hengl et al. 2017) | | Soil layer depth that reaches the bedrock |
| 5 | Occurrence_R_Horizon | WoSIS | | Probability of occurrence of R horizon |
| 6 | nbedrock | CLM5 simulation | | Soil layer number that reaches the bedrock |
| 7 | Koppen_Climate_2018 | (Beck et al. 2018) | Climate | Koppen Climate Classification |
| 8 | BIO1 | (Fick and Hijmans 2017) | | Annual Mean Temperature |
| 9 | BIO2 | | | Mean Diurnal Range |
| 10 | BIO3 | | | Isothermality |
| 11 | BIO4 | | | Temperature Seasonality |
| 12 | BIO5 | | | Max Temperature of Warmest Month |
| 13 | BIO6 | | | Min Temperature of Coldest Month |
| 14 | BIO7 | | | Temperature Annual Range |
| 15 | BIO8 | | | Mean Temperature of Wettest Quarter |
| 16 | BIO9 | | | Mean Temperature of Driest Quarter |
| 17 | BIO10 | | | Mean Temperature of Warmest Quarter |
| 18 | BIO11 | | | Mean Temperature of Coldest Quarter |
| 19 | BIO12 | | | Annual Precipitation |
| 20 | BIO13 | | | Precipitation of Wettest Month |
| 21 | BIO14 | | | Precipitation of Driest Month |
| 22 | BIO15 | | | Precipitation Seasonality |
| 23 | BIO16 | | | Precipitation of Wettest Quarter |
| 24 | BIO17 | | | Precipitation of Driest Quarter |
| 25 | BIO18 | | | Precipitation of Warmest Quarter |
| 26 | BIO19 | | | Precipitation of Coldest Quarter |
| 27 | USDA_Suborder | | | USDA 2014 Suborder Classes |
| 28 | WRB_Subgroup | | | WRB 2006 Subgroup Classes |
| 29 | Coarse_Fragments_v_0cm | | | |
| 30 | Coarse_Fragments_v_30cm | | | Coarse Fragments Volumetric |
| 31 | Coarse_Fragments_v_100cm | | | |
| 32 | Clay_Content_0cm | | | |
| 33 | Clay_Content_30cm | | | Clay Content |
| 34 | Clay_Content_100cm | | | |
| 35 | Silt_Content_0cm | | | |
| 36 | Silt_Content_30cm | (Hengl et al. 2017) | Soil Texture | Silt Content |
| 37 | Silt_Content_100cm | | | |
| 38 | Texture_USDA_0cm | | | |
| 39 | Texture_USDA_30cm | | | Texture Classes |
| 40 | Texture_USDA_100cm | | | |
| 41 | Sand_Content_0cm | | | |
| 42 | Sand_Content_30cm | | | Sand Content |
| 43 | Sand_Content_100cm | | | |
| 44 | Bulk_Density_0cm | | | |
| 45 | Bulk_Density_30cm | | | Bulk Density |
| 46 | Bulk_Density_100cm | | | |
| 47 | SWC_v_Wilting_Point_0cm | | | |
| 48 | SWC_v_Wilting_Point_30cm | | | Soil Water Capacity |
| 49 | SWC_v_Wilting_Point_100cm | | | |
| 50 | pH_Water_0cm | | | |
| 51 | pH_Water_30cm | | | Soil pH in H2O |
| 52 | pH_Water_100cm | (Hengl et al. 2017) | Soil Chemical Properties | |
| 53 | CEC_0cm | | | |
| 54 | CEC_30cm | | | Cation Exchange Capacity |
| 55 | CEC_100cm | | | |
| 56 | Garde_Acid | | | Grade of a Sub-Soil Being Acid |
| 57 | ESA_Land_Cover | ESA. Land Cover CCI Product User Guide Version 2. Tech. Rep. (2017) | | ESA Land Cover |
| 58 | cesm2_npp | | Vegetation | NPP |
| 59 | cesm2_npp_std | CLM5 simulation | | Standard deviation of NPP |
| 60 | cesm2_vegc | | | Vegetation Carbon Stock |

**Supplementary Table 2:** 21 biogeochemical parameters in CLM5

| No. | Name | Matrix Term | Corresponding Mechanism | Description | Unit | Prior Range | Reference |
|---|---|---|---|---|---|---|---|
| 1 | fl1s1 | | | Transfer fraction, from metabolic litter to fast SOC | unitless | [0.1, 0.8] | |
| 2 | fl2s1 | | | Transfer fraction, from cellulose litter to fast SOC | unitless | [0.2, 0.8] | |
| 3 | fl3s2 | | | Transfer fraction, from lignin litter to slow SOC | unitless | [0.2, 0.8] | |
| 4 | fs1s2 | | | Transfer fraction, from fast SOC to slow SOC | unitless | [0.0001, 0.4] | |
| 5 | fs1s3 | A | Microbial carbon use efficiency (CUE) | Transfer fraction, from fast SOC to passive SOC | unitless | [0.0001, 0.1] | Lawrence et al. 2019; Shi et al. 2018 |
| 6 | fs2s1 | | | Transfer fraction, from slow SOC to fast SOC | unitless | [0.1, 0.74] | |
| 7 | fs2s3 | | | Transfer fraction, from slow SOC to passive SOC | unitless | [0.0001, 0.1] | |
| 8 | fs3s1 | | | Transfer fraction, from passive SOC to fast SOC | unitless | [0.0001, 0.9] | |
| 9 | fcwdl2 | | | Transfer fraction, from coarse woody debris to cellulose litter | unitless | [0.5, 1] | |
| 10 | tau4cwd | | | Turnover time of coarse woody debris | year | [1, 6] | |
| 11 | tau4l1 | | | Turnover time of metabolic litter | year | [0.0001, 0.11] | Zhang et al. 2008 |
| 12 | tau4l2 | K | Substrate decomposability | Turnover time of cellulose litter | year | [0.1, 0.3] | |
| 13 | tau4s1 | | | Turnover time of fast SOC | year | [0.0001, 0.5] | |
| 14 | tau4s2 | | | Turnover time of slow SOC | year | [1, 10] | Xu et al. 2016 |
| 15 | tau4s3 | | | Turnover time of passive SOC | year | [20, 400] | |
| 16 | q10 | | | Temperature sensitivity | unitless | [1.2, 3] | Davidson et al. 2005; Davidson et al. 2006 |
| 17 | efolding | ξ | Environmental modifiers | E-folding parameter to calculate depth scalar | meter | [0.1, 1] | Lawrence et al. 2019; Shi et al. 2018 |
| 18 | w_scaling | | | Scaling factor to soil water scalar | unitless | [0.0001, 5] | |
| 19 | bio | V | Vertical transport | Bioturbation rate | m2/yr | [$3\times10^{-5}$ $16\times10^{-4}$] | Koven et al. 2015 |
| 20 | cryo | | | Cryoturbation rate | m2/yr | [$3\times10^{-5}$ $5\times10^{-4}$] | |
| 21 | beta | I | Carbon input | Vertical distribution of carbon input | unitless | [0.5, 0.9999] | Lawrence et al. 2019; Shi et al. 2018 |

**Supplementary Table 3:** Sensitivity Analysis of BINN Modeling Choices

| Modelling Part | Test Choices | Test NSE | Training Stability |
|---|---|---|---|
| Original Settings | Smooth L1 + Leaky ReLU + Sigmoid | 0.611 ± 0.036 | - |
| Loss Function | L1 Loss | 0.589 ± 0.045 | - |
| | L2 Loss | 0.615 ± 0.025 | Extreme Para |
| | Para Squared Diff | 0.616 ± 0.042 | Extreme Para |
| Activation | ReLU | 0.608 ± 0.038 | - |
| | Hard Sigmoid | 0.59 ± 0.047 | - |
| Prior Range | Prior Range Increased by 10% | 0.6 ± 0.04 | - |
| | Prior Range Decreased by 10% | 0.602 ± 0.039 | - |

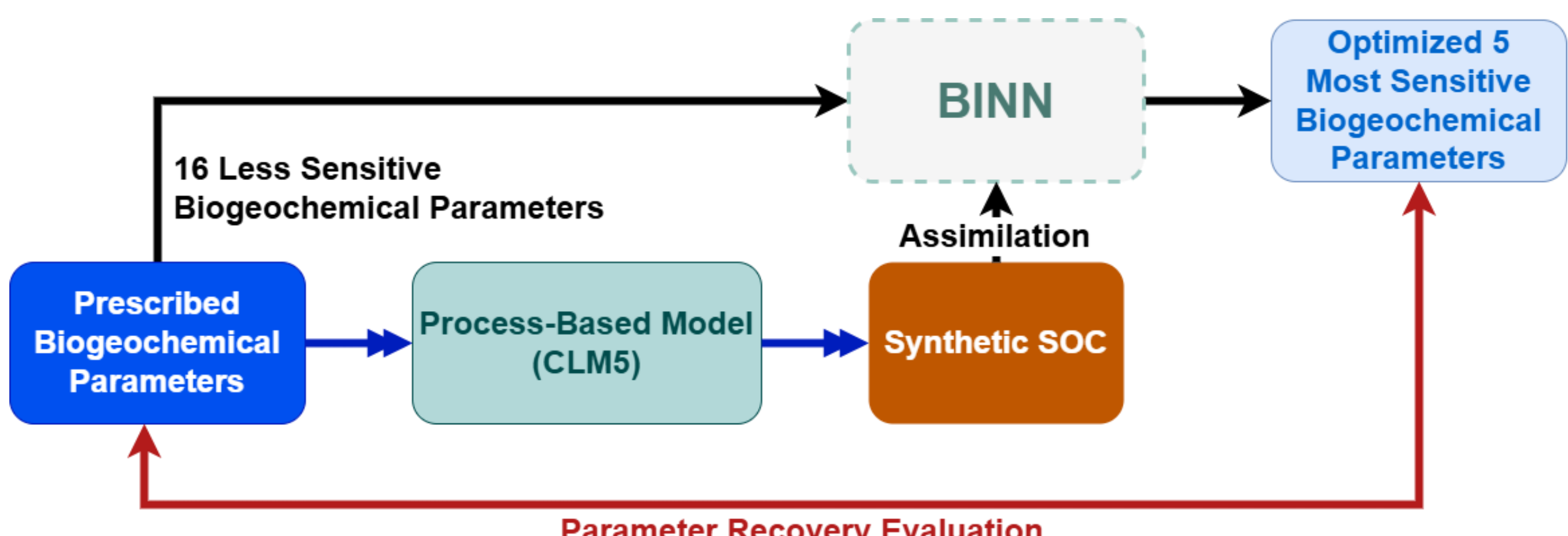

**Supplementary Figure 2: Schematic of the parameter recovery experiment to evaluate BINN's ability to retrieve the model parameters.** The parameter recovery experiment involves three main steps. 1). Blue two-headed arrows: Synthesizing a SOC dataset using CLM5 with prescribed parameter values (21 parameters). 2). Single-headed arrows: Using the synthetic SOC dataset to train BINN to predict the 5 most sensitive parameters. 3). Red double arrow: By comparing the BINN-predicted parameters with the prescribed parameters used to generate the synthetic dataset, we can assess BINN's effectiveness in retrieving the parameters of processes regulating SOC from observational data.

**Supplementary Table 4:** Eight environmental forcings for CLM5

| Variable Names | Description | Resolution |
|---|---|---|
| nbedrock | Soil layer number that reaches the bedrock | 0.5 degree, average monthly values from the monthly record of 20-year simulation after the system reaches the steady state |
| ALTMAX | Maximum active layer depth of current year | |
| ALTMAX_LASTYEAR | Maximum active layer depth of last year | |
| CELLSAND | Sand content | |
| NPP | Net primary productivity | |
| SOILPSI | Soil water potential | |
| TSOI | Soil temperature | |
| O_SCALAR | Oxygen scalar for decomposition | |
| FPI_vr | Nitrogen scalar for decomposition | |

(a)

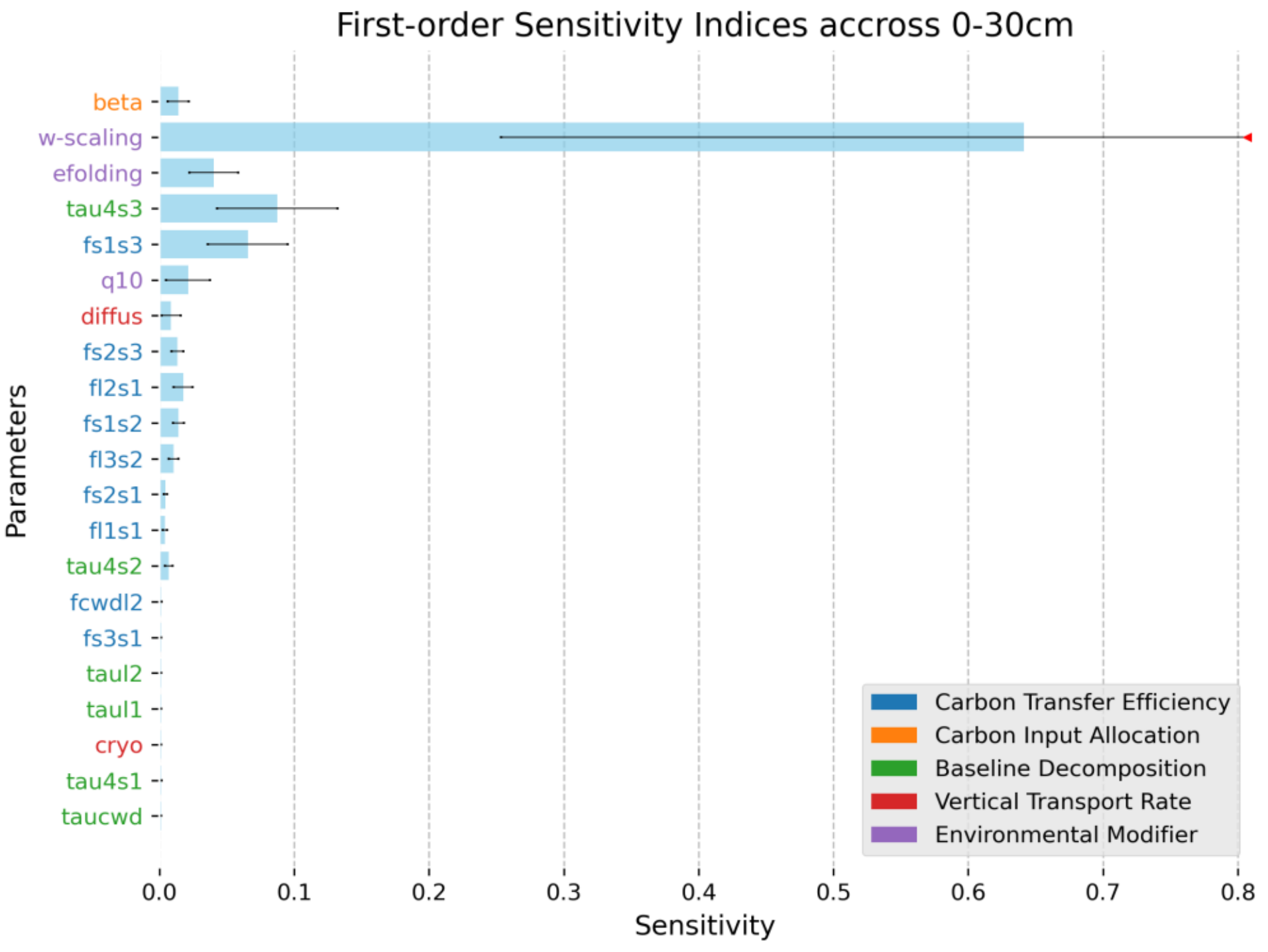

(b)

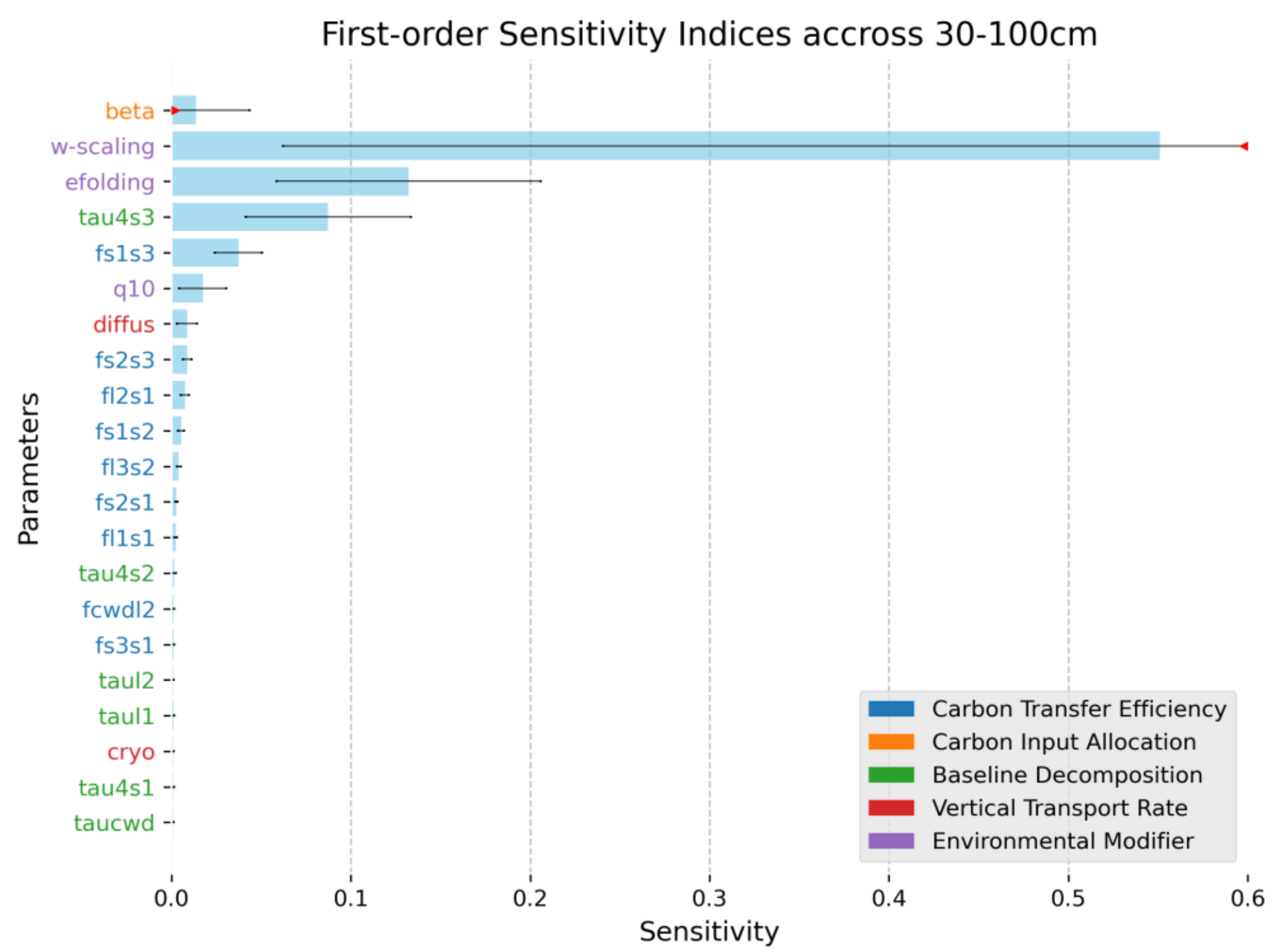

(c)

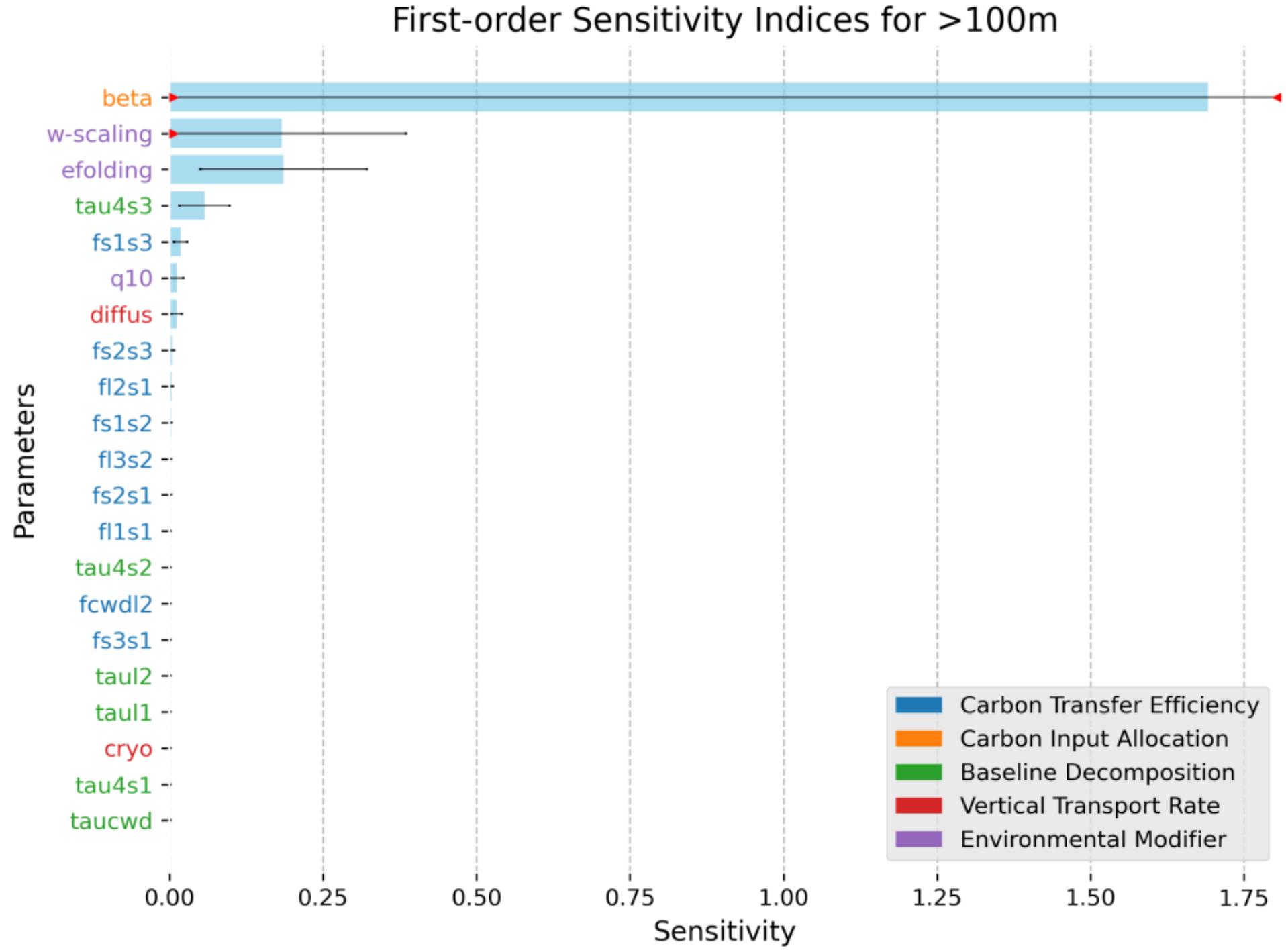

**Supplementary Figure 3:** Sensitivity indices for CLM5 biogeochemical parameters across: (a) 0-30cm, (b) 30-100 cm, and (c) >100 cm.

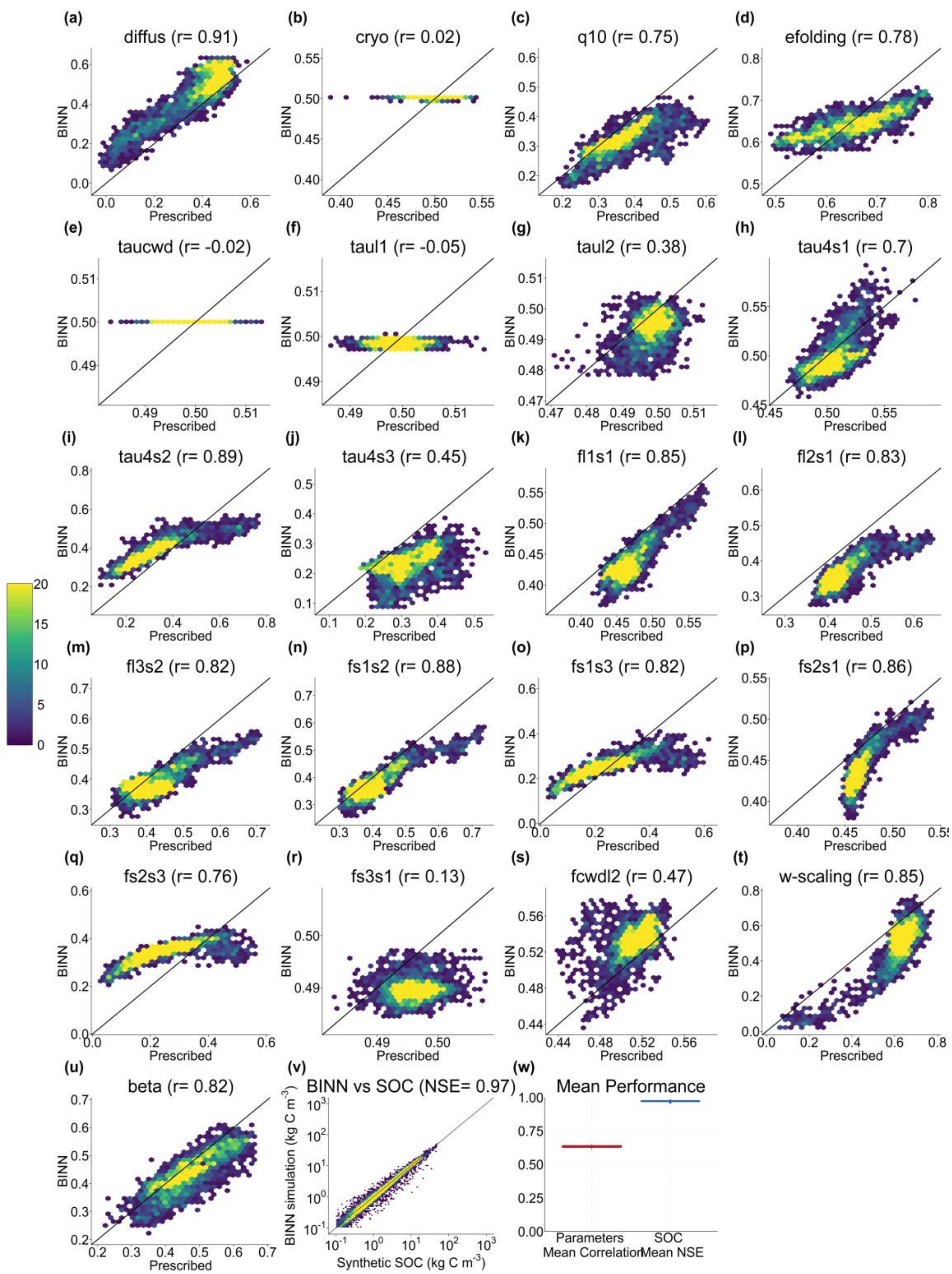

**Supplementary Figure 4:** BINN parameter recovery for all CLM5 parameters. (a–u) Scatter plots of BINN-predicted versus prescribed values for each parameter. (v) Density scatter of BINN simulated SOC versus synthetic SOC. (w) Summary across a 10-fold cross-validation: mean correlation for all parameters (predicted vs. prescribed) and NSE for SOC simulations.

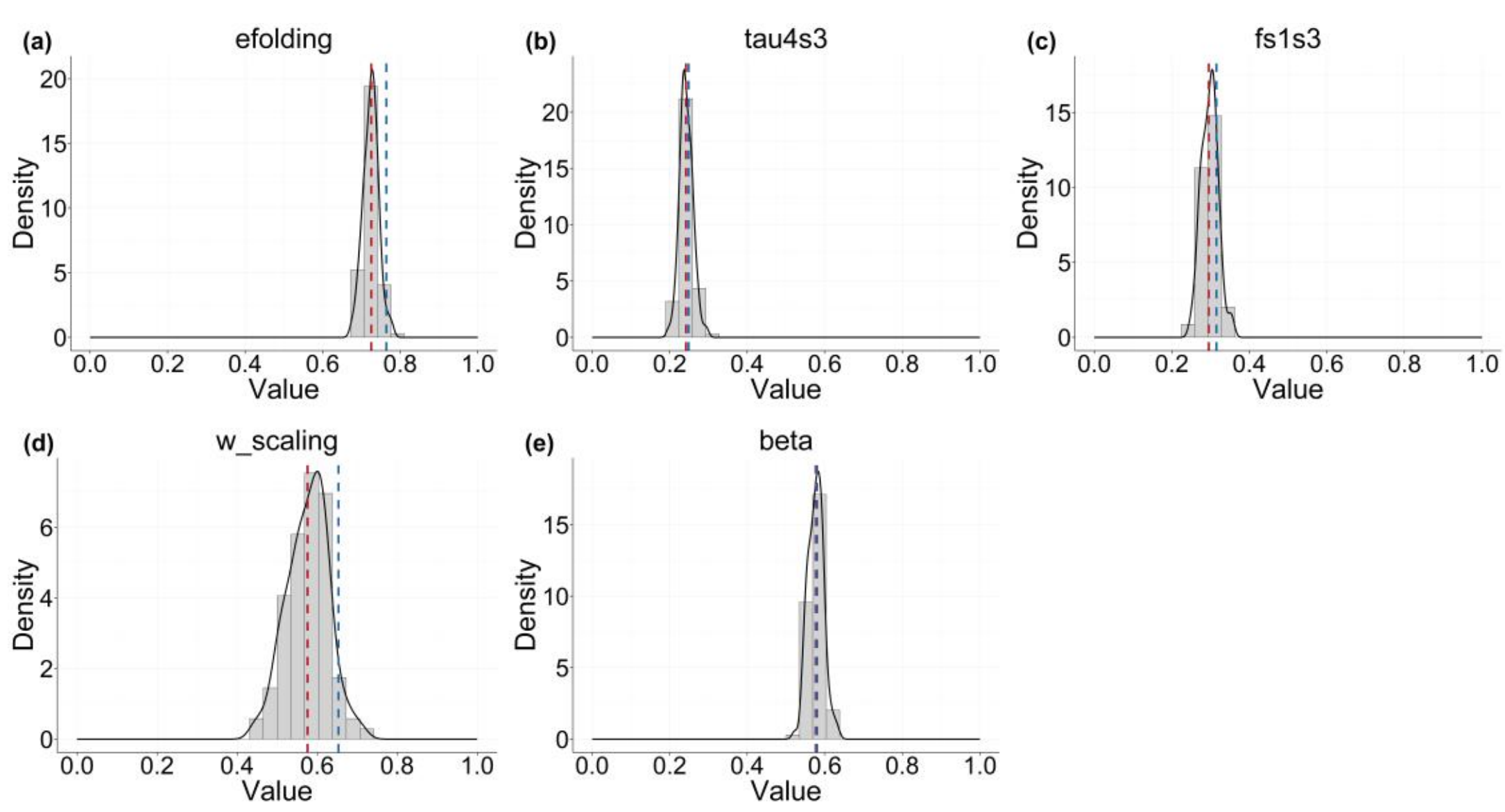

**Supplementary Figure 5: Uncertainty in posterior distributions of BINN predicted parameters in relation to Monte Carlo dropout in the parameter recovery experiment at one site.** For each parameter, the marginal posterior is shown, with the vertical blue and red dashed lines indicating the prescribed ("true") and BINN's point estimate, respectively.

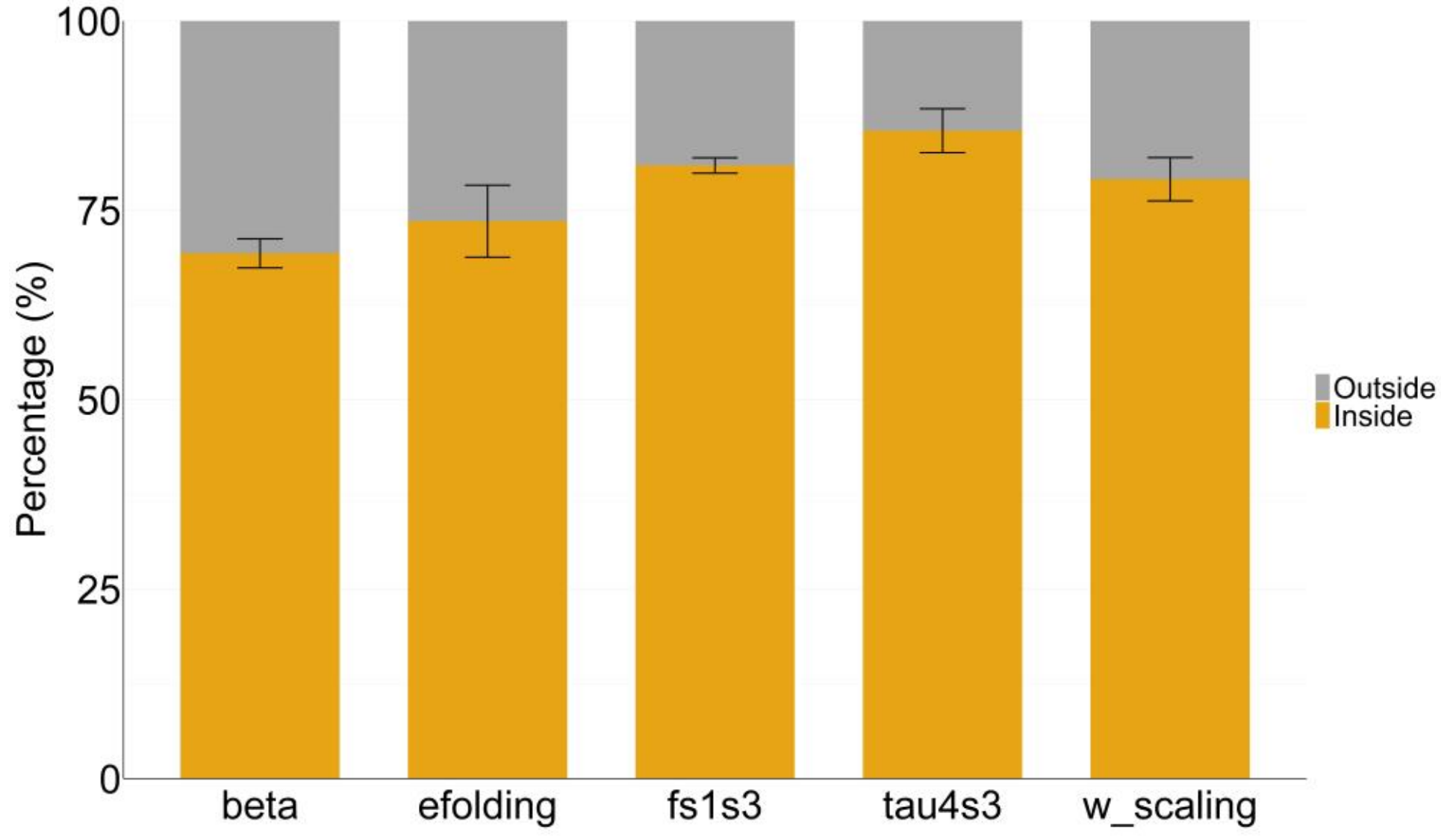

**Supplementary Figure 6: Coverage of parameters' posterior ranges of BINN on the prescribed parameters.** The coverage was quantified by the percentage of prescribed ("true") parameter values that fall into the Monte Carlo–dropout posterior range across test sites. Error bars show the mean ± SD across the 10 cross-validation folds of the synthetic parameter-recovery experiment.

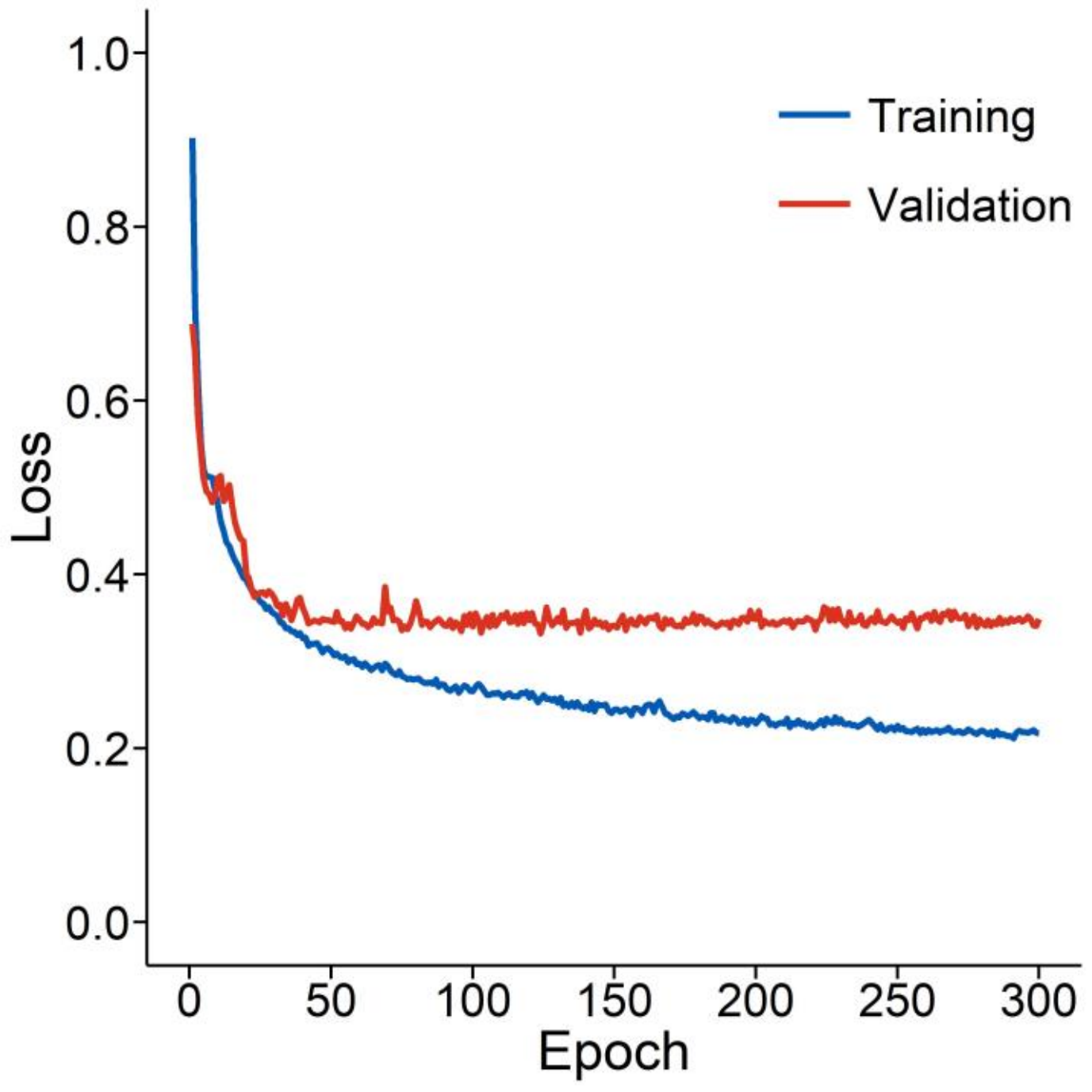

**Supplementary Figure 7:** Training and validation inefficiency history for one cross-validation fold with median NSE value.

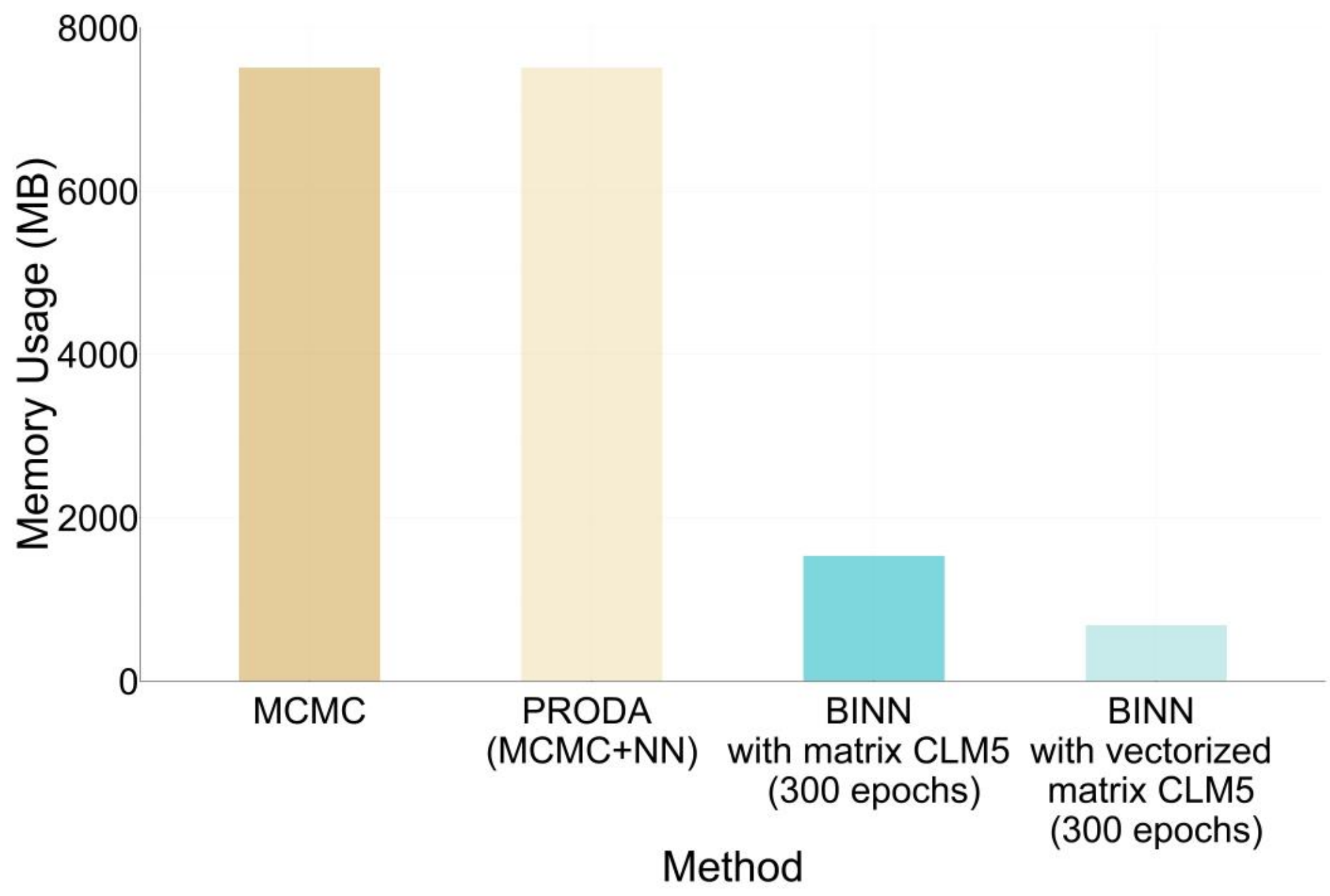

**Supplementary Figure 8:** Comparative analysis of the peak memory (peak resident set size, RSS) required for integrating 2,000 soil profiles into the process-based model (CLM5).

**Supplementary Table 5:** Comparison among different data assimilation methods

| Criteria | BINN | PRODA | MCMC | Kalman Filter | Genetic Algorithm |
|---|---|---|---|---|---|
| Optimization Speed | Fast | Slow | Slow | Fast | Slow |
| Optimization Target | Parameters | Parameters | Parameters | States | Parameters |
| Recognition of Spatial/Temporal Heterogeneity | Yes | Yes | No | No | No |
| Uncertainty Assessment | No | No | Yes | Yes | No |
| Multisource Data | Yes | Yes | Yes | Yes | Yes |
| Key Refs | This study | Tao et al. 2020 Frontiers[2] | Hararuk et al. 2014 JGR[3] | Williams et al. 2005 GCB[4] | Barrett et al. 2002 GBC[5] |